\documentclass[5p,twocolumn]{elsarticle} % two col publish format
\usepackage{amsfonts}
\usepackage{hyperref}
\usepackage{siunitx}
\usepackage{amsmath}
\usepackage{graphicx}
\usepackage{color}
\usepackage{xcolor}
\usepackage{float}

\newcommand{\pfrac}[2]{\frac{\partial #1}{\partial #2}}

\definecolor{ref1}{rgb}{0.0, 0.0, 0.0}
\definecolor{ref2}{rgb}{0.0, 0.0, 0.0}
\definecolor{ref3}{rgb}{0.0, 0.0, 0.0}

\newcommand{\rfone}[1]{\textcolor{ref1}{#1}}
\newcommand{\rftwo}[1]{\textcolor{ref2}{#1}}
\newcommand{\rfthree}[1]{\textcolor{ref3}{#1}}

\journal{Journal of High Energy Density Physics}

\begin{document}

\title{Cross-Code verification and sensitivity analysis to effectively model the electrothermal instability}

\author[VT,LLNL]{R. L. Masti}
\ead{rlm7819@vt.edu}
\author[LLNL]{C. L. Ellison}
\author[TECHX]{J. R. King}
\author[TECHX]{P. H. Stoltz}
\author[VT]{B. Srinivasan \corref{cor1}}
\ead{srinbhu@vt.edu}

\cortext[cor1]{Corresponding Author}
\address[VT]{Virginia Polytechnic Institute and State University, Blacksburg, VA 24060, USA}
\address[LLNL]{Lawrence Livermore National Laboratory, Livermore, CA 94550, USA}
\address[TECHX]{Tech-X Corporation, 5621 Arapahoe Ave., Boulder, CO 80303, USA}

\date{\today}

\begin{abstract}
This manuscript presents verification cases that are developed to study the electrothermal instability (ETI).  Specific verification cases are included to ensure that the unit physics components necessary to model the ETI are accurate, providing a path for fluid-based codes to effectively simulate ETI in the linear and nonlinear growth regimes.  Two software frameworks with different algorithmic approaches are compared for accuracy in their ability to simulate diffusion of a magnetic field, linear growth of the ETI, and a fully nonlinear ETI evolution.  The nonlinear ETI simulations show early time agreement, with some differences emerging, as noted in the wavenumber spectrum, late into the nonlinear development of ETI. A sensitivity study explores the role of equation-of-state (EOS), vacuum density, and vacuum resistivity. EOS and vacuum resistivity are found to be the most critical factors in the modeling of nonlinear ETI development.
\end{abstract}


\maketitle

\section{Introduction}\label{sec:introduction}
  The current-driven electrothermal instability (ETI) forms when the material resistivity is dependent on temperature, occurring in nearly all Z-pinch-like high energy density (HED) platforms.\cite{oreshkin2008} Previous work models the early time behavior of current-driven metallic explosions for pulsed wire array configurations as well as for imploding metal liner configurations such as in the magnetized liner inertial fusion (MagLIF) experiments.\cite{peterson2012, slutz2010} A number of codes have been used to simulate and understand the ETI, making it important to quantify how numerical modeling choices influence the evolution of the instability. \citep{oreshkin2008, peterson2012, peterson2013, peterson2014}
  
  This work provides a series of verification cases in both linear and nonlinear regimes ensuring ETI-relevant unit physics is simulated accurately. Comparing these cases across codes highlights which differences between the codes are most important when simulating ETI such as time integration schemes for diffusion, spatial differencing methods, and numerical treatment of the highly resistive vacuum in which the ETI target resides. Additionally, this work performs cross-code comparisons for simulations of nonlinear ETI in regimes relevant to MagLIF and other pulsed-power driven HED platforms.  

  The two codes are USim, a commercially available multiphysics fluid code from Tech-X \cite{USim}, and Ares, a Lawrence Livermore National Laboratory (LLNL) multiphysics radiation-hydrodynamics code.\cite{morgan2016, ellison2018, darlington2001} USim is an unstructured-mesh-based Eulerian code while Ares is a structured-mesh-based arbitrary Lagrangian-Eulerian (ALE) code, and for this study, diffusion is temporally handled explicitly in USim and implicitly in Ares. Both codes solve the resistive magneto-hydrodynamics (MHD) equations with thermal conductivity. Obtaining similar results with such different algorithmic approaches provides confidence in the underlying discretization techniques and implementation in both codes.

  This paper is structured as follows. Section~\ref{sec:code} presents code descriptions for Ares and USim along with details on equation-of-state (EOS). Section~\ref{sec:comp} presents code verification that ensures the magnetic diffusion is captured accurately relative to an analytic solution and that the linear growth of ETI, in regimes of relevance to the Z-machine experiments \cite{peterson2012, peterson2013, peterson2014}, compares well with theory. Following code verification in the analytic and linear ETI regime, Section~\ref{sec:nonlineti} presents comparisons of nonlinear ETI including a sensitivity study of nonlinear ETI dynamics to EOS treatment, vacuum resistivity, and vacuum density.

  The key contributions of this paper are two-fold.  First, analytic and theory-driven unit physics cases provide verification of the critical physics components necessary to accurately model ETI. Second, the nonlinear studies highlight the sensitivity of vacuum parameters and EOS in converged nonlinear ETI behavior. The sensitivity analysis shows nonlinear ETI behavior is influenced by vacuum resistivity more than vacuum density. 

\section{Code Descriptions}\label{sec:code}
For this study, the Ares and USim codes solve the magnetohydrodynamic equations which are given in conservative form as

\begin{equation}\label{eqn:cont}
  \pfrac{\rho}{t} + \nabla \cdot \left[\rho \mathbf{u}\right] = 0,
\end{equation}
\begin{equation}\label{eqn:mtm}
  \pfrac{\rho \mathbf{u}}{t} + \nabla \cdot \left[\rho \mathbf{u}\mathbf{u}^T - \frac{\mathbf{B}\mathbf{B}^T}{\mu_0} + \mathbb{I}\left(P + \frac{|\mathbf{B}|^2}{2\mu_0}\right)\right] = 0,
\end{equation}
\begin{equation}\label{eqn:ener}
  \pfrac{\epsilon}{t} + \nabla \cdot \left[\left(\epsilon + P + \frac{|\mathbf{B}|^2}{2\mu_0}\right)\mathbf{u}- \frac{1}{\mu_0}\mathbf{B}\cdot\mathbf{u}\mathbf{B}\right] = S_{\epsilon},
\end{equation}
and 
\begin{equation}\label{eqn:magf}
  \pfrac{\mathbf{B}}{t} = -\nabla \times S_\mathbf{E},
\end{equation}
where $\mu_0$, $\rho$, $\mathbf{u}$, $P$, $\epsilon$ and $\mathbf{B}$, are the magnetic permeability of free space, mass density, 3D velocity, total pressure, total energy density, and magnetic field vector, respectively. $S_\epsilon$ and $S_\mathbf{B}$ represent the source terms for ohmic heating and thermal conduction in the former, and resistive magnetic diffusion in the latter. %\footnote{This could be changed for a relative permeability of a non vacuum region such as in the liner where $\mu_0\rightarrow\mu_r\mu_0$ where $\mu_r$ can be taken from the magnetic susceptibility} 

The codes differ in multiple ways such as in their variable storage scheme (e.g. zone or node storage), mesh evolution, diffusion evaluation, and diffusion time integration. 
The different treatment of the diffusion terms, represented in $S_\epsilon$ and $S_B$ in Equations~\ref{eqn:ener}~and~\ref{eqn:magf}, respectively, significantly influences ETI growth. For this work, $S_\epsilon$ is given by
\begin{equation}\label{eqn:sepsusim}
  S_\epsilon = \frac{-1}{\mu_0}\nabla \cdot (\frac{\eta}{\mu_0}\nabla\times\mathbf{B}), 
\end{equation}
where $\eta$ is the electrical resistivity, and the contribution to the $S_\mathbf{B}$ in Equation~\ref{eqn:magf} is given by 
\begin{equation}\label{eqn:smagusim}
  S_\mathbf{B} = -\nabla\times(\frac{\eta}{\mu_0}\nabla\times\mathbf{B}).
\end{equation}
For this study, differences exist algorithmically in the temporal integration and spatial differentiation of these terms. Section~\ref{sec:nonlineti} explores these differences leading to substantially different computational challenges and numerical limitations.

\subsection{USim Code}\label{sec:code:usim}
USim uses finite-volume algorithms on an unstructued Eulerian grid to solve conservative equation systems. In the simulations presented here, USim uses the Monotone Upwinding Scheme for Conservation Laws (MUSCL) to perform cell interface reconstruction for the computation of the Eulerian fluxes.\cite{van1979} \rftwo{For these fluxes, USim utilizes the Harten-Lax-van Leer-Discontinuities (HLLD) approximate Riemann solver given by \citet{miyoshi2005} (5-wave) modified to incorporate the changes to the waves introduced by the real-gas EOS}.\cite{king2020} Additionally, USim utilizes hyperbolic divergence cleaning as given by \citet{dedner2002}. For hyperbolic temporal integration, USim uses a 2$^{nd}$ order Runge-Kutta time integration scheme with variable time step.

For this study, USim uses super-time-stepping (STS) to handle the parabolic terms embedded in $S_\epsilon$ and $S_\mathbf{B}$. STS modifies the number of Runge-Kutta stages for the parabolic terms so they can be evolved at the hyperbolic time step. \cite{alexiades1996} The number of stages is proportional to the ratio of the hyperbolic time step to the diffusion time step, which can be many orders of magnitude for the nonlinear ETI simulations in Section~\ref{sec:nonlineti}, albeit there are limits to the number of stages before STS begins to impact accuracy.\cite{alexiades1996} Although USim's STS is capable of 2$^{nd}$ order accuracy, only the first-order accurate implementation is used for direct comparisons between codes. Section~\ref{sec:comp:magdiff} tests the magnetic diffusion contribution to $S_\epsilon$ in Equation~\ref{eqn:ener} which is given by Equation~\ref{eqn:sepsusim}, and contribution to $S_\mathbf{B}$ in Equation~\ref{eqn:magf} which is given by Equation~\ref{eqn:smagusim}.

USim implements the divergence and curl operators through a polynomial fit approximation as this algorithm is suitable for problems with a general unstructured mesh. With this method a field variable, such as $B_x$, is fitted with a multi-dimensional polynomial and the value of the derivative is computed to obtain the differentiated quantity; e.g. the current density.\cite{mavriplis2003} Discontinuities in field variables can cause oscillations in the fitting procedure of the least-squares method. Circumventing these oscillations requires using a large stencil which reduces or smooths out the magnitude of the derivatives at the discontinuity.\footnote{A stencil of 20 was found to be sufficient in evaluating diffusive fluxes} As an example, $S_\mathbf{B}$ from Equation~\ref{eqn:smagusim}, experiences large gradients due to the discontinuous resistivity at the vacuum-liner interface during the nonlinear ETI simulations, and a small stencil would over represent the gradient magnitude at this interface. While USim typically uses a 2nd order derivative reconstruction, the large gradients in the diffusive quantities in Eq. 6 require a first order reconstruction due to numerical instability with the higher order polynomial fits.

%\footnote{The least-squares method in USim is particularly important when using unstructured meshes; however, this work only employed structured meshes.}

Nonlinear ETI simulations in Section~\ref{sec:nonlineti} include thermal diffusion in addition to magnetic diffusion augmenting the $S_\epsilon$ term. USim evolves the total energy density given in Equation~\ref{eqn:ener}, so an inverse EOS operation is performed at every time step to get the temperature from the internal energy density. For this study, USim applies thermal diffusion through a source term given as
%\footnote{The inverse EOS operation described here is going from density and specific internal energy density to temperature
\begin{equation}\label{eqn:thermdiff}
  \pfrac{T}{t} = \nabla\cdot(\alpha\nabla T), 
\end{equation}
where $\alpha$ is the thermal diffusivity. This equation requires the use of an additional EOS operation to compute internal energy density ($\epsilon_{int}$) updating the total energy density ($\epsilon$) from Equation~\ref{eqn:ener} through the relation $\epsilon=\epsilon_{int}+1/2\rho \mathbf{u}^2+1/(2\mu_0)\mathbf{B}^2$. For the nonlinear ETI simulations in Section~\ref{sec:nonlineti}, both codes use thermal conduction, as it improves numerical stability and is physically relevant in the linear ETI growth phase of the simulation.\cite{oreshkin2008,peterson2012} 

\rfone{USim handles the multi-material setup of the nonlinear ETI simulation (liner-vacuum) through the use of a marker which is a unit-less identifier (-1 to 1). This marker is evolved with the normalized advective fluxes of Equation~\ref{eqn:cont} which follows the movement of each material respectively. Due to the density voids created in the nonlinear ETI growth and the lack of vacuum energy conservation (see Section~\ref{sec:nonlineti} and Figure~\ref{fig:fft}), this marker is filtered such that a zone containing liner material above the interface cannot transition to a vacuum zone (marker is always $>0$ above interface). This implementation does not allow for ejection of liner material into the vacuum, but it does allow for seldom transition from liner material to vacuum along the continuous interface through the marker going from $>0$ to $<0$.}

%Adding thermal diffusion is important to the nonlinear ETI simulations in Section~\ref{sec:nonlineti} because it improves the numerical stability, and it's a known physically stabilizing mechanism against ETI growth.\cite{oreshkin2008,peterson2012}

%the growth rate goes like - k_z^2 \kappa_therm, which means that the thermal conduction smooths higher k modes much more than it smooths lower k modes. This is a good signature for something that improves the numerics. I suggest acknowleding that aspect of the dissipation, as well as making it clear that thermal conduction is included in both codes.

\subsection{Ares Code}\label{sec:code:ares}

Ares is one of LLNL's multiphysics radiation hydrodynamics codes specializing in inertial confinement fusion (ICF), high energy density (HED) physics, and energetic materials \cite{darlington2001, morgan2016, ellison2018}. At its core, Ares solves single-fluid multi-material multi-component\footnote{I.e., Ares evolves a single fluid velocity but multiple material densities and temperatures within any multi-material zones, and each material allows multiple components (equations of state) that are required to be in pressure and temperature equilibrium with other components of the same material in the zone.} Euler or Navier-Stokes hydrodynamic equations on a structured, arbitrary Lagrangian-Eulerian (ALE), adaptive mesh refinement (AMR) grid. Depending on the application, additional physics packages are incorporated in an operator split fashion. \rfthree{Major physics packages include resistive and extended magnetohydrodynamics, laser ray tracing and energy deposition, single- or multi-group radiation diffusion, discrete ordinates ($S_n$) radiation transport\cite{lathrop1968,castor2004}, Reynolds-Averaged Navier-Stokes (RANS) turbulence models, and thermonuclear burn.}
  
For this work, all of the Ares simulations use the 2D resistive MHD package without AMR. The 2D MHD package assumes that currents reside in the $x-y$ or $r-z$ simulation plane, while a single component of the magnetic field evolves perpendicular to the simulation plane. During the Lagrange step, the zone-centered magnetic field is frozen into the fluid. \rfthree{After the Lagrange step, the mesh can optionally be relaxed towards its initial position according to the user's ALE prescription.} All mesh variables are then interpolated from the post-Lagrange mesh to the relaxed mesh using conservative, finite-volume, total variation diminishing flux-limited advection schemes. In the case of the magnetic field, the finite volume advection preserves magnetic fluxes. Note that for the purposes of this work, Ares was run in full relaxation ``Eulerian mode", which compares nicely to USim's Eulerian formulation.

Resistive diffusion of the magnetic field and the hydrodynamics motion are treated separately using operator splitting methods. Both the magnetic diffusion equation and the thermal diffusion are advanced implicitly in time using a first-order accurate backward-Euler method. Similarly, both diffusion operations employ a second-order accurate finite volume spatial discretization.\cite{pert1981} This method is akin to a bilinear finite element discretization. 

Ohmic heating is applied explicitly in time after the implicit magnetic diffusion update. The updated magnetic field is differenced to calculate edge-centered currents according to Ampere's Law. The ohmic heating incurred by the edge-centered currents is partitioned into the two adjacent zones by treating the two zones as resistors in parallel. 

\rfone{Ares handles multi-material dynamics with a volume-of-fluid approach. This approach assigns a volume fraction to each material present within a given zone. In addition to the sub-zonal volume, each material is allowed its own sub-zonal thermodynamic state including density, temperature, and pressure. However, only a single (node-centered) fluid velocity is maintained (thus the single-fluid multi-material designation for the code). For MHD, a zone-averaged conductivity is required for magnetic diffusion and ohmic heating. For this study, Ares uses a mass-fraction-weighted average of the conductivities for zones that contain multiple materials.}

\subsection{Equation of State}\label{sec:code:eos}

Since ETI growth depends on the resistivity of a material, and the resistivity is a function of the material's state, an accurate EOS is important. HED simulations of experiments often rely on tabular EOS libraries to provide accurate representations of the material state across a wide range of densities and temperatures. These EOS tables provide the $P$ and the $\epsilon$ as functions of density and temperature, including into HED regimes. In previous work\cite{peterson2012, peterson2013}, the SESAME EOS database was used to model ETI specifically using SESAME 3720 (SES3720) \cite{SESAME3720} for an aluminum EOS, and Sandia Lee-More based Desjarlais (QLMD) tables for aluminum transport properties.\cite{desjarlais2002}

The nonlinear ETI simulations in Section~\ref{sec:nonlineti} employ an analytic Birch-Murnaghan EOS (BMEOS) for ease of code-code comparisons. The magnetic diffusion and linear ETI simulations in Section~\ref{sec:comp} employ an ideal gas EOS. Sensitivity studies in Section~\ref{sec:nonlineti:eos} assess how the nonlinear ETI behavior differs between BMEOS and SES3720 EOS. For this study, the QLMD effective ionization table is used in conjunction with the BMEOS to span a large state space. BMEOS is an analytic equation of state determined through data regression, and this work uses the functional form used by \citet{mcbride2015} where the pressure and internal energy are given by \cite{birch1947, murnaghan1937, murnaghan1944, mcbride2015}
\begin{equation}\label{eqn:bmp}
  \begin{split}
    P = P_0 + \frac{3}{2} A_1 \left[\left(\frac{\rho}{\rho_0}\right)^{g_1} - \left(\frac{\rho}{\rho_0}\right)^{g_2}\right]\\
    \left[1+ \frac{3}{4}(A_2-4)\left[\left(\frac{\rho}{\rho_0}\right)^g - 1\right]\right],
  \end{split}
\end{equation}
where for aluminum $P_0 = (1+Z_\text{eff})k_B \rho T/m$, $m=\SI{4.509e-26}{\kilo\gram}$, $A_1=\SI{76e9}{\pascal}$, $\rho_0=\SI{2700}{\kilo\gram\per\cubic\metre}$, $g_1=\SI{7/3}{}$, $g_2=\SI{5/3}{}$, $A_2=\SI{3.9}{}$, and $g=\SI{2/3}{}$ with $Z_\text{eff}$, $k_B$, $\rho$, $T$, representing the effective ionization level, Boltzmann constant, density $[\si{\kilo\gram\per\cubic\metre}]$, and temperature $[\si{\kelvin}]$, respectively.\cite{mcbride2015} Similarly, the specific internal energy density is given by 
\begin{equation}\label{eqn:bme}
  \begin{split}
    \epsilon = \epsilon_0 + \frac{9}{16} A_1 \rho_0^{-1} \Bigg[A_2\left[\left(\frac{\rho}{\rho_0}\right)^g - 1\right]^3 + \\ \left[\left(\frac{\rho}{\rho_0}\right)^g - 1\right]^2\left[6 - 4\left(\frac{\rho}{\rho_0}\right)^g\right]\Bigg],
  \end{split}
\end{equation}
where $\epsilon_0 = 3/2 (1+Z_\text{eff}) k_B T/m$. BMEOS has a max difference to SES3720 of 40\%; see \ref{app:bmeos}.

Implementing tabulated EOS or tabulated transport coefficients requires a choice of interpolation algorithm such as bilinear or bicubic. Section~\ref{sec:nonlineti:eos} shows the effect of interpolation algorithm and EOS on the Ares nonlinear ETI simulation. For this study, Ares uses LLNL's LEOS \cite{more1988, young1995, fritsch2003} algorithms for table interpolation, and USim uses Los Alamos's EOSPAC interpolation library. \cite{cranfill1983} 

\section{Cross Code Verification}\label{sec:comp}

This work provides a guide to running nonlinear ETI simulations by sequentially verifying the individual physics components relevant to ETI. The first verification test is of magnetic diffusion where solutions from the two codes are compared against an analytical result. The second verfication test is a linear ETI simulation with negligible magnetic diffusion relative to the ohmic heating in $S_\epsilon$ of Equation~\ref{eqn:ener}, isolating the ohmic heating and resistive feedback mechanisms.  

\subsection{Magnetic Diffusion}\label{sec:comp:magdiff}

This test case involves an $x$-directed magnetic field varying sinusoidally along the $y$ direction resistively diffusing due to a constant resistivity in space and time. This test uses Cartesian coordinates with the fluid initially at rest. The magnetic field diffuses towards the steady state solution of a constant field. Comparing this time evolution to the analytically-derived solution quantifies the numerical error. 

The diffusion equation in Equations~\ref{eqn:magf}~and~\ref{eqn:smagusim} with constant resistivity reduces the curl operations to a simple Laplacian diffusion equation in Cartesian coordinates. In order to isolate magnetic diffusion from the full MHD equation, the initial conditions and strength of electrical resistivity must satisfy certain conditions. These conditions are that any thermal pressure due to ohmic heating be negligble relative to the magnetic pressure, and that any motion due to the magnetic pressure occurs at much longer time-scales than the magnetic diffusion time-scale. Using a plasma beta, the ratio of thermal pressure to magnetic pressure, of unity satisfies the first condition, and a Lundquist number of unity satisfies the second condition. \rftwo{Although only the magnetic field is needed to analyze Equation~\ref{eqn:smagusim}, the codes evolve the full MHD equations; hence, a low plasma beta and a low Lundquist number are chosen to study the isolated effect of magnetic diffusion in Equations~\ref{eqn:magf}~and~\ref{eqn:smagusim}}.

The chosen simulation grid uses an $x$ domain of \SI{0.25}{\metre} and an $y$ domain of \SI{1}{\metre} with a resolution of 50x200 grid cells, respectively. The uniform initial state is $P_0=$\SI{1.0133e5}{\pascal} and $\rho_0=$\SI{0.164}{\kilogram\per\cubic\metre} with an ideal gas equation of state (EOS) using $\gamma=\frac{5}{3}$.\footnote{The choice of initial state or EOS has no impact on this test} The initial magnetic field is $\mathbf{B_0}=\left<\SI{0.5044}{}\cos(2\pi y), 0 ,0\right>\si{\tesla}$, and the initial electrical resistivity is $\eta=$\SI{1.396e-3}{\ohm\metre}. Given these parameters, the characteristic magnetic diffusion rate is 
\begin{equation}\label{eqn:mdgr}
  \gamma_{md} = \frac{4 \pi^2}{L_y L_u}\sqrt{\frac{2 P_0}{\beta\rho}}\approx\SI{4.387e4}{\per\second}, 
\end{equation}
and the chosen simulation end time is $t_f=$\SI{45.59}{\micro\second} ($2/\gamma$). 

\begin{figure}[h]
  \centering
  \includegraphics[width=1.0\linewidth]{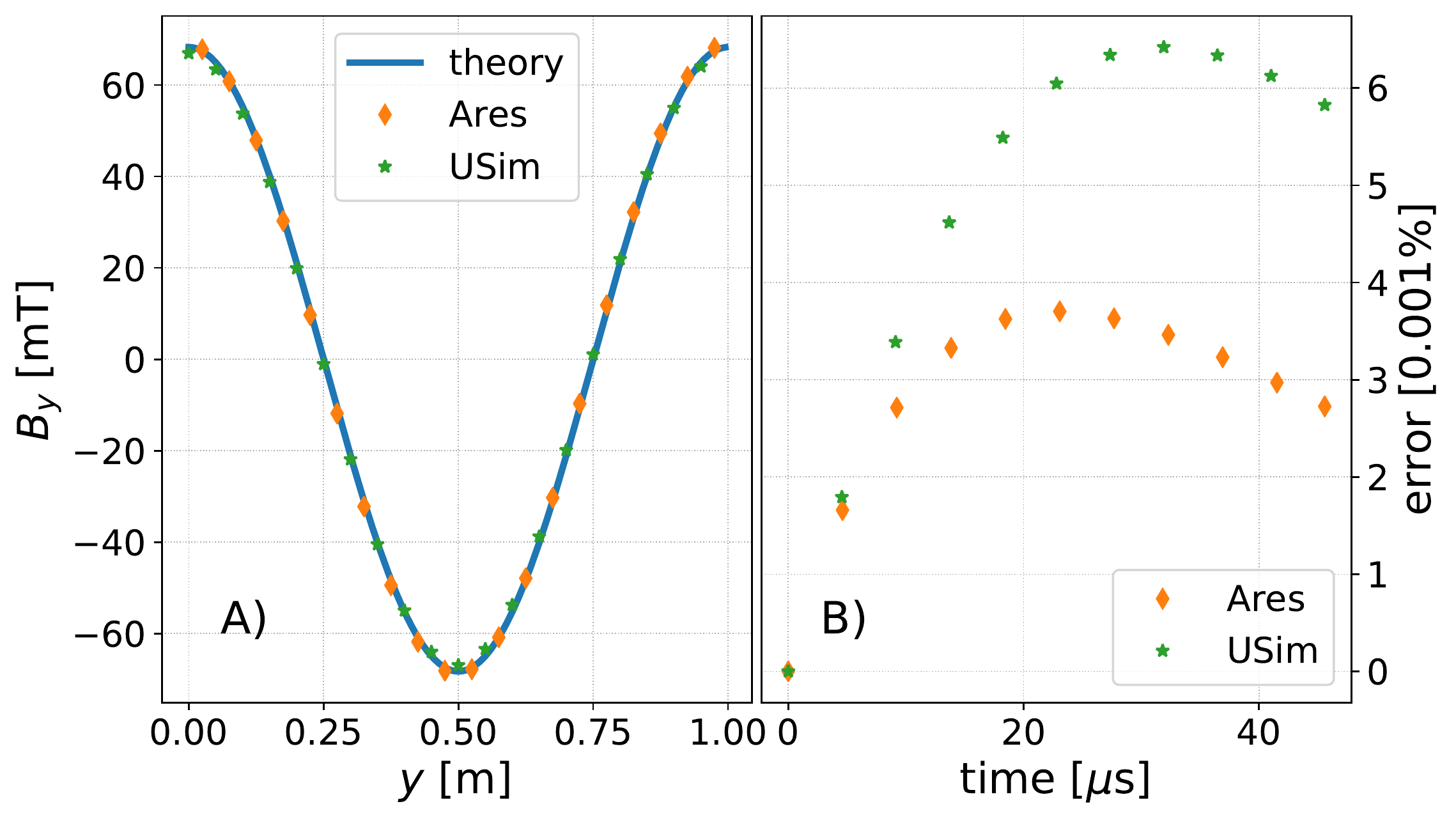}
  \caption{Plot A on the left shows the $x$-direction magnetic field in [\si{\milli\tesla}] along the vertical direction at the magnetic diffusion simulation end time of \SI{45.59}{\micro\second} over a subset of the simulation range. Plot B on the right shows the L$_2$ norm of the error between the simulated magnetic field and the analytically-derived magnetic field over time.} \label{fig:md}
\end{figure}

Figure~\ref{fig:md} presents the error of both codes as a function of time, and an instantaneous lineout of each simulation along with the analytically-derived solution. This configuration results in a maximum global error of less than 0.01\% for both the Ares and USim simulations. This low error provides confidence in the magnetic diffusion capabilities of both codes, and is critical for resolving the nonlinear ETI magnetic diffusion wave.

\subsection{Linear ETI}\label{sec:comp:lineti}
\begin{figure}[h]
  \centering
  \includegraphics[width=1.0\linewidth]{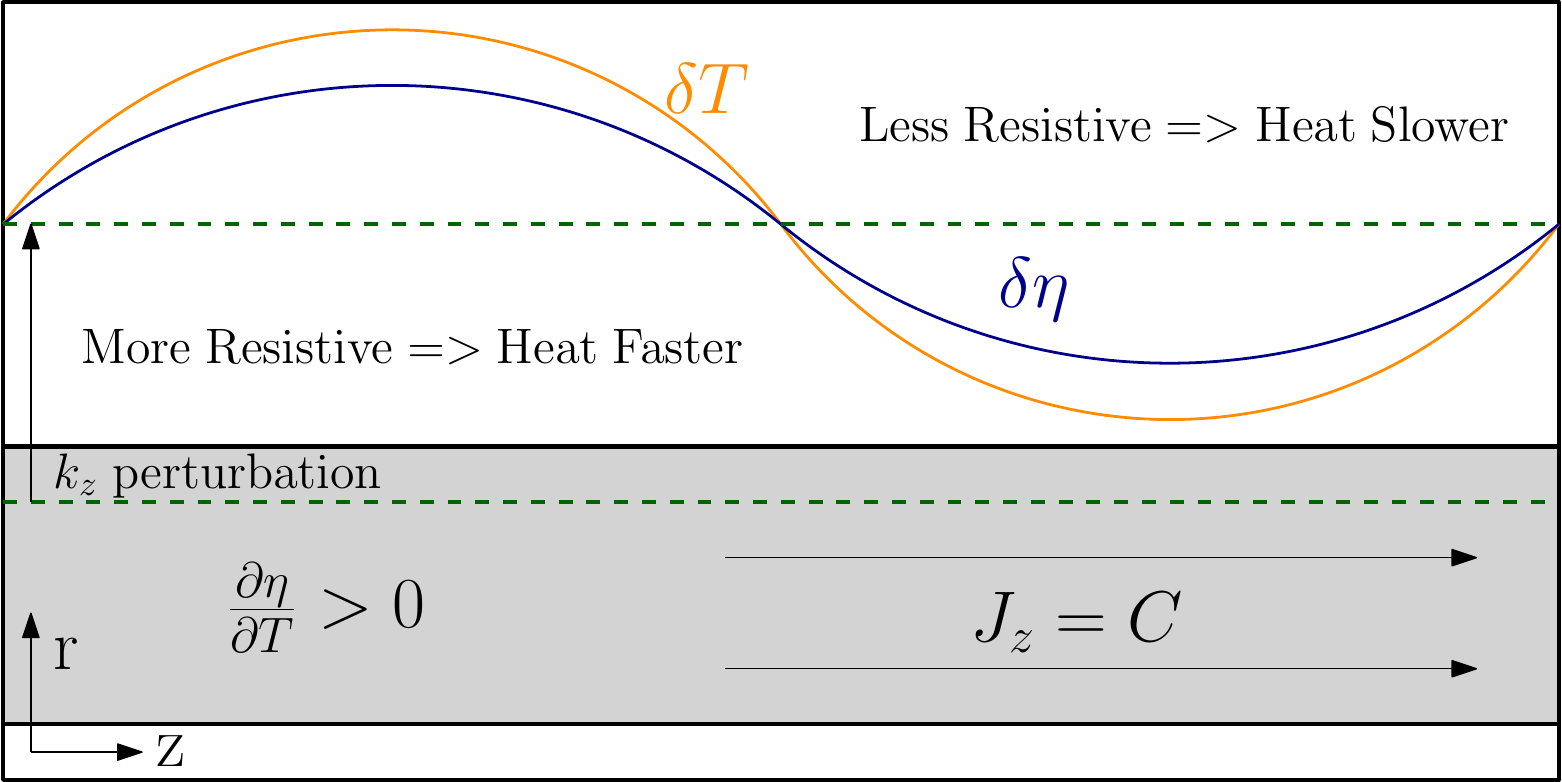}
  \caption{Schematic depicting the linear ETI test in cylindrical coordinates wherein a spatially varying resistivity exists inside a uniformly distributed current. The perturbed temperature (or the internal energy density) perturbs resistivity, and provided the resistivity increases with temperature, this configuration is ETI unstable as implied by Equation~\ref{eqn:etigr}.} \label{fig:etilin}
\end{figure}

ETI occurs whenever ohmic heating is applied to a material with a temperature-dependent resistivity. The combination of the changing resistivity and ohmic heating creates a positive feedback loop causing hot spots to develop internally. The linear ETI growth rate is given by\cite{ryutov2000, peterson2012}
\begin{equation}\label{eqn:etigrfull}
  \gamma = \frac{\eta_T J_z^2\left(1 - \frac{2\cos^2\alpha}{1+\gamma/\gamma_0}\right) - k^2 \kappa}{\rho \epsilon_T}, 
\end{equation}
\rfthree{where $\gamma_0  = 2k\eta/\mu_0 \Delta r$, and where $\mathbf{k}$, $\kappa$, $\eta$, $T$, $\epsilon_T$, and $\alpha$, are the wavevector, the thermal conductivity, the resistivity, the temperature, the partial derivative of the specific internal energy (\si{\joule\per\kilogram}) with respect to temperature, and the angle between the wavevector and magnetic field, respectively}. For $\eta_T \equiv \pfrac{\eta}{T} > 0$ (most commonly the case for metals at solid densities and low temperatures), the maximum growth occurs when $\alpha=\SI{90}{\degree}$, resulting in a growth rate of
\begin{equation}\label{eqn:etigr}
  \gamma = \frac{\eta_T J_z^2 - k_z^2 \kappa}{\rho \epsilon_T}\;.
\end{equation}
While the magnetic diffusion test verifies the effect of the electrical resistivity on the magnetic field evolution, this linear ETI test verifies the effect of the electrical resistivity on the internal energy density evolution. Figure~\ref{fig:etilin} shows the problem setup. In the absence of thermal conductivity, the growth rate becomes 
\begin{equation}\label{eqn:sgr}
  \gamma=\frac{\eta_T J_z^2}{\rho \epsilon_T}.
\end{equation}
\rftwo{This form of the theoretical growth rate depends heavily on ohmic heating, so reproducing this analytical result through simulation provides confidence in each code's ability to capture ohmic heating and the feedback of such heating on the evolution of the material state}. For this simulation, the initial parameters (relevant current, length, and time scales) are set to reproduce ohmic heating in a typical pulsed power regime. This test uses aluminum as the conducting material following the state parameters and conductivities derived from the SES3720 and QLMD 29373 tables, respectively.\cite{SESAME3720, desjarlais2002}.

This test uses initial parameters of $\rho_0 = \SI{2700}{\kilogram\per\cubic\metre}$, $T_0=\SI{250}{\kelvin}$, $I_0 = \SI{10}{\mega\ampere}$, over an annulus with a thickness of $\SI{500}{\micro\metre}$ starting at a radial location of \SI{2.68}{\milli\metre}. Uniformally distributing the current in the annulus results in a current density of $J_z = \SI{1.09e12}{\ampere\per\squared\metre}$, and is similar to the values from Figure~4 of \citet{peterson2013} (\SIrange{1e12}{7e12}{\ampere\per\squared\metre}). From the growth rate defined in Equation~\ref{eqn:sgr}, the $\eta_T$ and the $\epsilon_T$ for this simulation use values consistent with realistic solid metallic parameters relevant to pulsed power HED regimes. 

\begin{figure}[h]
  \centering
  \includegraphics[width=1.0\columnwidth]{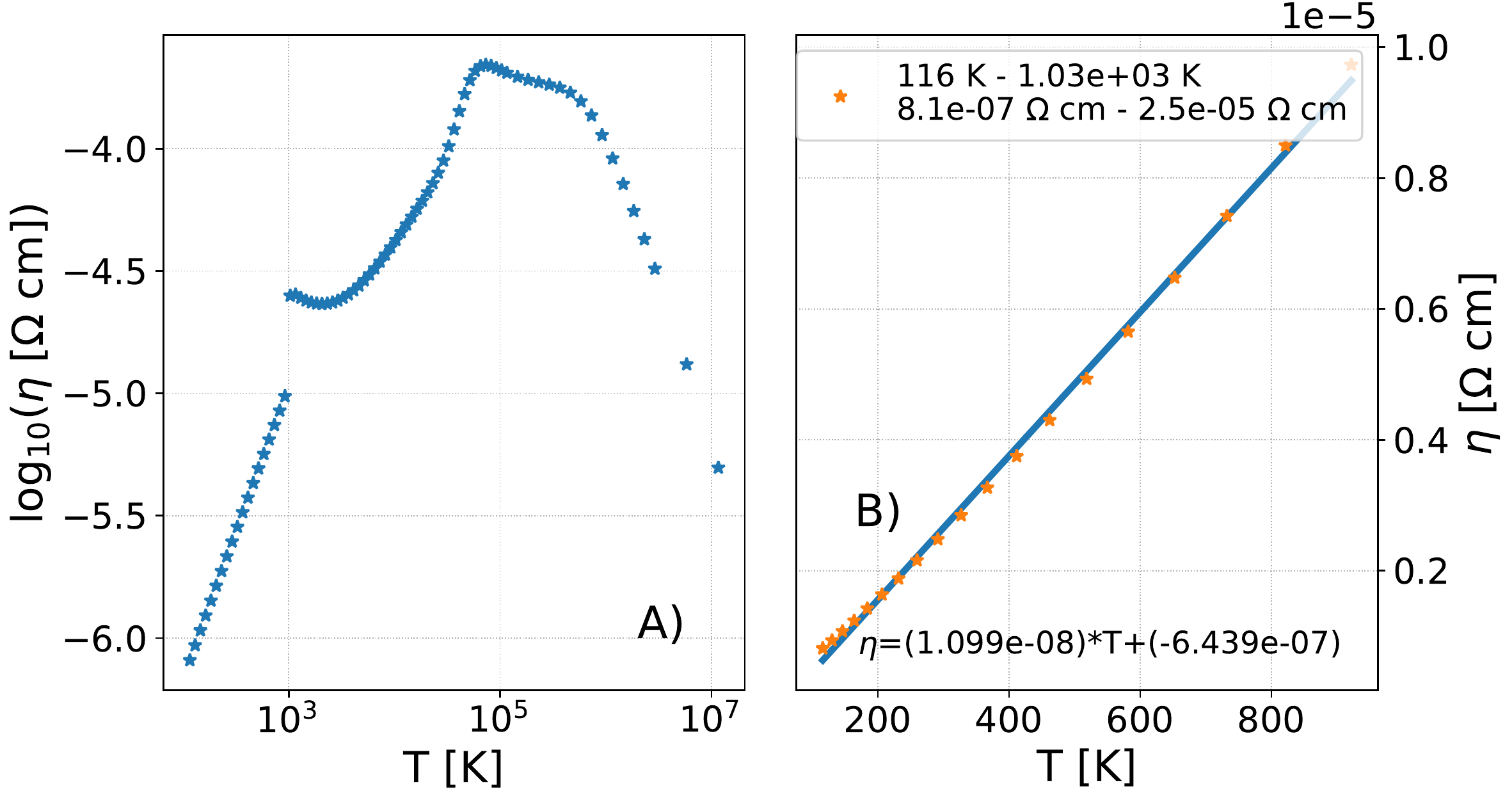}
  \caption{Plot A on the left shows a constant density contour of the logged QLMD conductivity table for aluminum for a certain temperature range.\cite{desjarlais2002} Plot B on the right shows a subset of Plot A in the low temperature regime, where the resistivity linearly increases with temperature. The fitted linear region in plot B yields an estimate for $\eta_T$.} \label{fig:eta}
\end{figure}

Figure~\ref{fig:eta} shows the conductivity for aluminum at solid density over the entire range of the QLMD table in panel A) and a linear fit to a small range of temperatures in panel B).\cite{desjarlais2002} The fit from plot B yields an $\eta_T=\SI{1.099e-8}{\second\per\kelvin}$ (in mks: $\eta_T=\SI{1.099e-10}{\ohm\metre\per\kelvin}$). This $\eta_T$ is valid for constant density solid aluminum between \SIrange{200}{900}{\kelvin}; note that the electrical resistivity is more sensitive to the density than the temperature in this state space region. \cite{peterson2012}

The specific heat capacity, $\epsilon_T$, is approximately \SI{822}{\joule\per\kilogram\per\kelvin} based on SES3720 at the initial state. This test uses an ideal gas EOS with the adiabatic index, $\gamma$, chosen to maintain a constant $\epsilon_T$. Knowing $\epsilon_T$, the adiabatic index is $\gamma = 1+\left[k_B/(m \epsilon_T)\right]$ where $k_B$ is the Boltzmann constant and $m$ is the atomic weight, resulting in an adiabatic index of $\gamma \approx \SI{1.373}{}$. 

\begin{figure}[!htb]
  \centering
  \includegraphics[width=1.0\columnwidth]{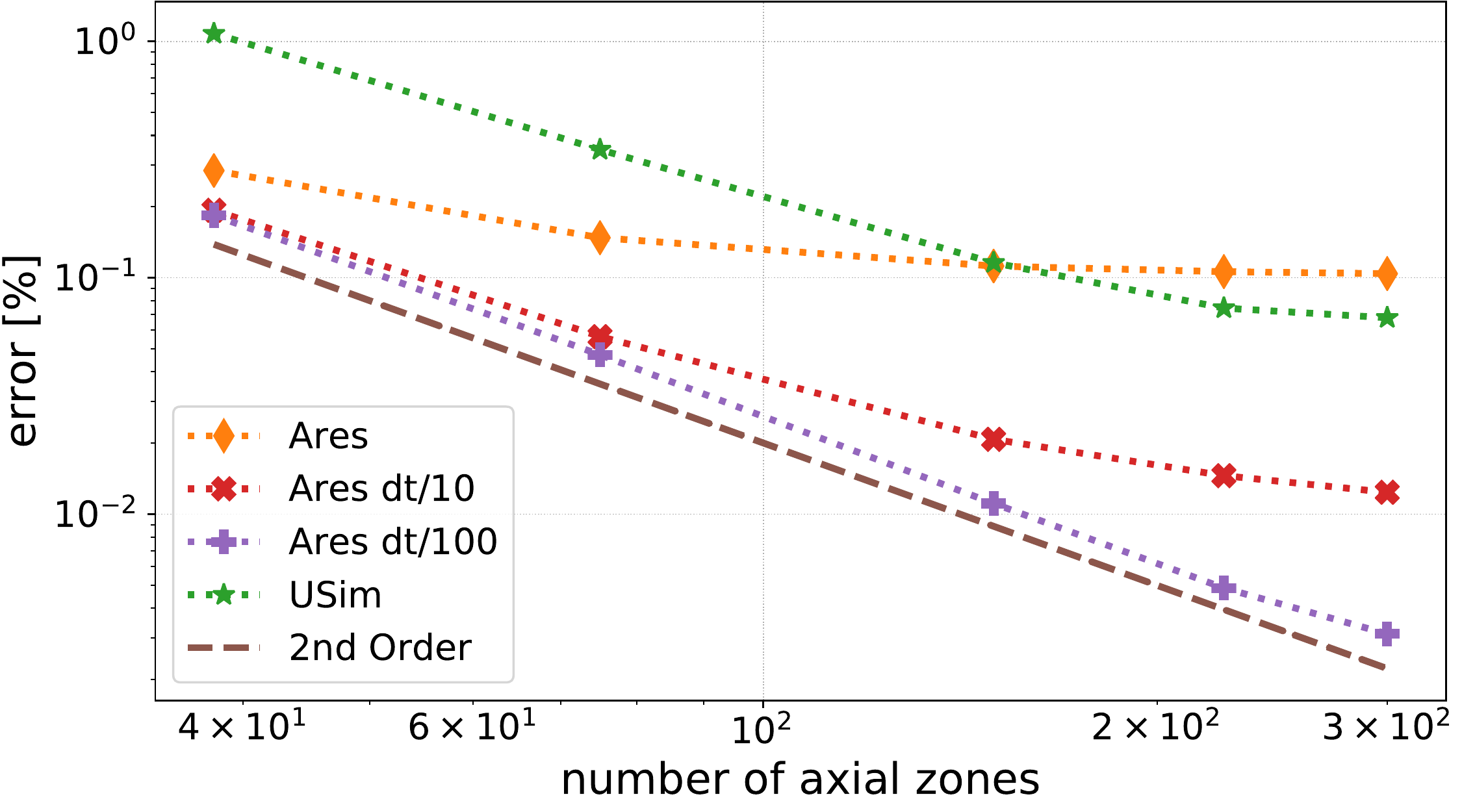}
  \caption{Plot of percent error, relative to the theoretical growth rate, of the linear ETI simulation in Section~\ref{sec:comp:lineti} for different radial and axial resolutions. The Ares convergence exhibits second-order \rftwo{spatial convergence (error $\propto \Delta z^{-2}$=zone size$^{-2}$)} provided the time discretization errors are small (USim was not tested in this limit), and both codes asymptote to similar percent error when using similar time steps .} \label{fig:etigrloglog}
\end{figure}

The simulation radial domain is from \SIrange{2.54}{3.16}{\milli\metre} with the inner annulus radius set to \SI{2.66}{\milli\metre}, and the simulation axial domain is $\pm\SI{.25}{\milli\metre}$ and is arbitrary based on Equation~\ref{eqn:sgr} (absent $k_z$ dependence). The resolution varies in the radial direction from \SIrange{38}{300}{} cells and in the axial direction from \SIrange{25}{200}{} using a factor of two refinement levels to produce the convergence plot in Figure~\ref{fig:etigrloglog}. 

With these parameters, the ETI growth rate from Equation~\ref{eqn:etigr} is $\gamma \approx \SI{5.891e7}{\per\second}$. The end time for this simulation is three growth periods corresponding to $3/\gamma\approx\SI{57.5}{\nano\second}$. The error between the simulated linear ETI growth\footnote{Obtained by fitting an exponential growth of the temperatue deviation from the mean (maximum - minimum)} and the theoretical growth rate produces Figure~\ref{fig:etigrloglog} using different spatial resolutions and time steps. 

\rftwo{The rate (order) of convergence describes how the error (difference between exact solution and numerical approximation) decreases for increasing spatial or temporal resolution}. Figure~\ref{fig:etigrloglog} presents convergence results of USim and Ares for the linear ETI simulation. For the largest time step, Ares asymptotes to 0.1 percent error and USim approaches first-order spatial convergence. This difference is due to Ares using a fixed time step and USim using an adaptive time step. The error depicted in Figure~\ref{fig:etigrloglog} has contributions from both the spatial discretization and temporal discretization. The temporal discretization error, at sufficiently small time step size, becomes small relative to spatial discretization error thereby recovering the Ares spatial order-of-accuracy of second order for uniform meshes. Comparing the asymptotic errors exhibited by Ares in the high resolution limit, there is reasonable agreement with the anticipated first-order accuracy in time outlined in Section~\ref{sec:code}. Additionally, while Ares uses a fixed maximum time step for this simulation, USim has an adaptively changing time step that decreases slightly as the solution evolves, which likely contributes to the lower asymptotic error.

In summary, both codes accurately capture the theoretical growth rate to within 0.1\% in the asymptotic limit given the maximum stable time step. Figure~\ref{fit:etigrloglog} shows the Ares spatial order-of-accuracy is second order, and the temporal order-of-accuracy is approximately first order. This case shows excellent agreement between the linear ETI growth rate from Equation~\ref{eqn:etigr} and the simulated linear ETI growth across both codes.

\section{Nonlinear ETI}\label{sec:nonlineti}

\begin{figure}%htb]
  \centering
  \includegraphics[width=0.9\columnwidth]{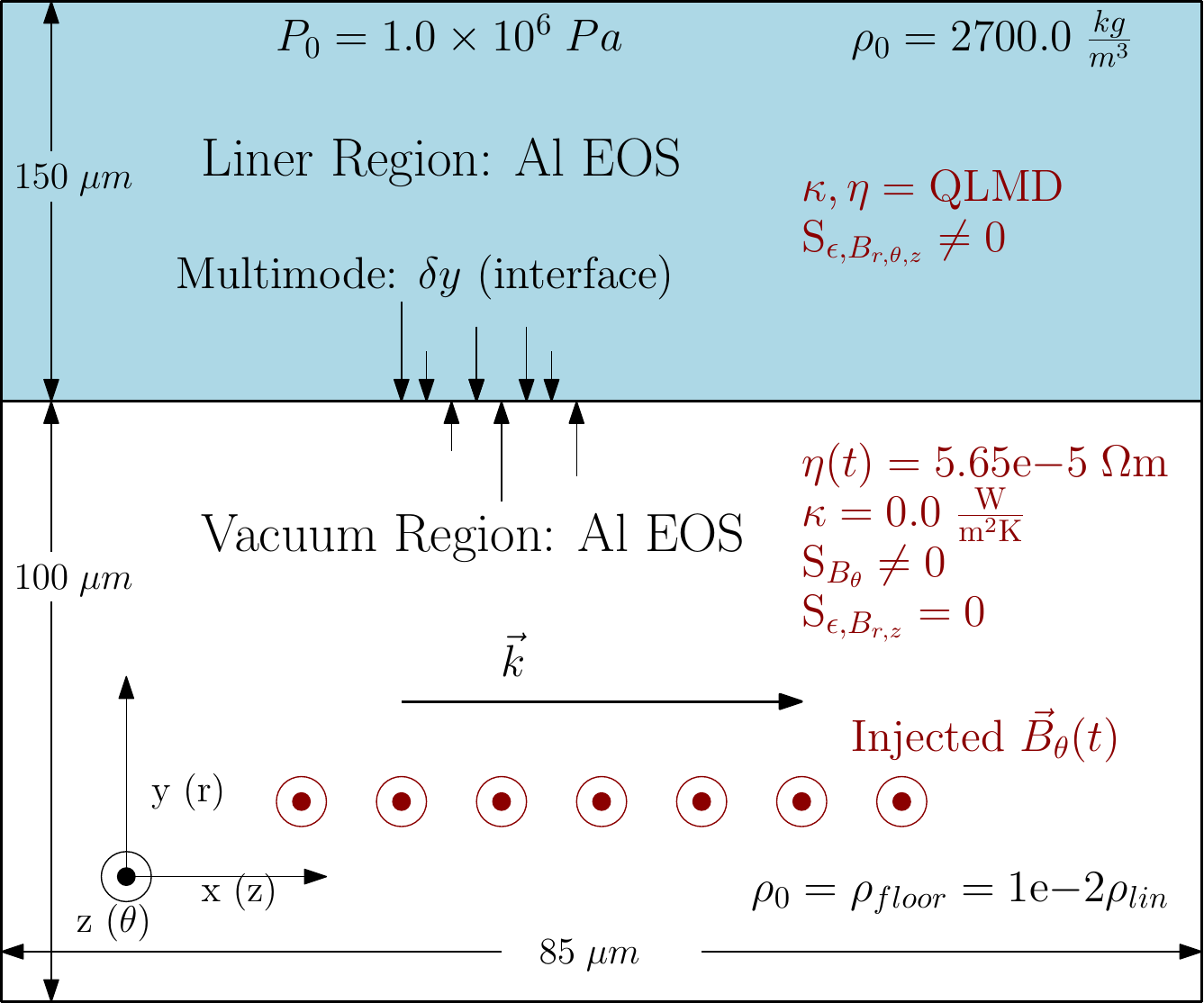}
  \caption{\rftwo{The simulation setup for the nonlinear ETI discussed in Section~\ref{sec:nonlineti}, showing the different regions, densities, source terms, and conductivities.}}\label{fig:simsetup}
\end{figure}

The nonlinear ETI case explores, in planar coordinates, the effect of resolving the magnetic diffusion wave through a medium (aluminum). The subsequent current redistribution, and the spatially non-uniform ohmic heating leads to nonlinear growth of ETI. This nonlinear ETI case builds upon the verfied magnetic diffusion from Section~\ref{sec:comp:magdiff} and the verified ohmic heating from Section~\ref{sec:comp:lineti} to run a physically-relevant ETI simulation for solid cylindrical and wire explosion regimes.\cite{oreshkin2008, peterson2012, peterson2013} This specific case, as shown in Figure~\ref{fig:simsetup}, is based on the simulation work done by \citet{peterson2013}. 

\subsection{Baseline Nonlinear ETI}\label{sec:nonlineti:base}

For the current source in these simulations, Ares allows the user to directly specify the current as a function of time whereas USim requires specifying the value of the magnetic field at the boundary. Within the code, Ares uses Ampere's law to specify the magnetic field at the boundary, and given the same coordinate system, this is the same as directly specifying the magnetic field at the boundary as done in USim. Limitations in the accuracy of USim's curl operator in cylindrical coordinates lead to using planar geometry in both codes for these simulations. Although the simulations are in planar geometry, the specified magnetic field boundary condition is consistent with the cylindrical geometry.

\begin{figure}%htb]
  \centering
  \includegraphics[width=1.0\columnwidth]{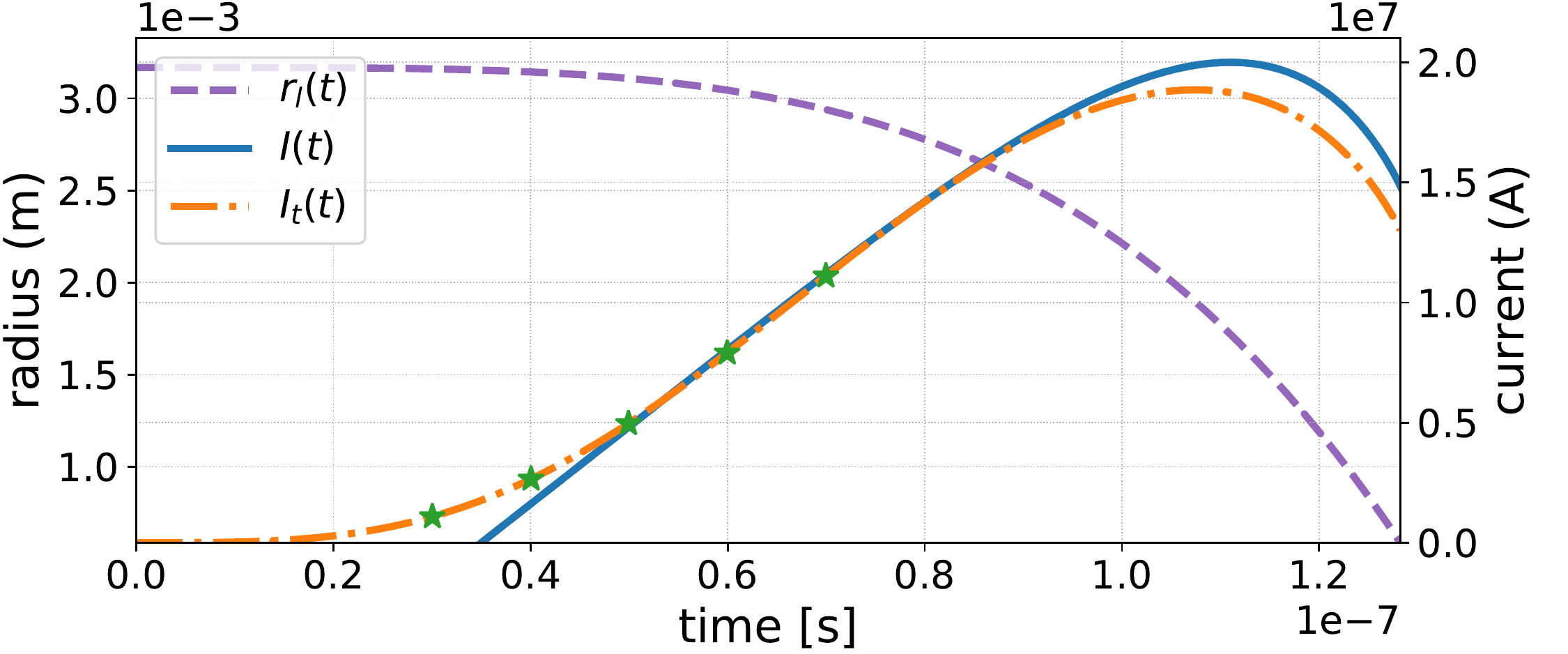}
  \caption{Liner outer radius $r_l(t)$, normal current $I(t)$, and adjusted current $I_t(t)$ as a function of time with the adjusted current profile providing inclusion of the prepulse phase typical of a Z-machine shot.}\label{fig:rAndI}
\end{figure}

\begin{figure*}%htb]
  \centering
  \includegraphics[width=1.0\linewidth]{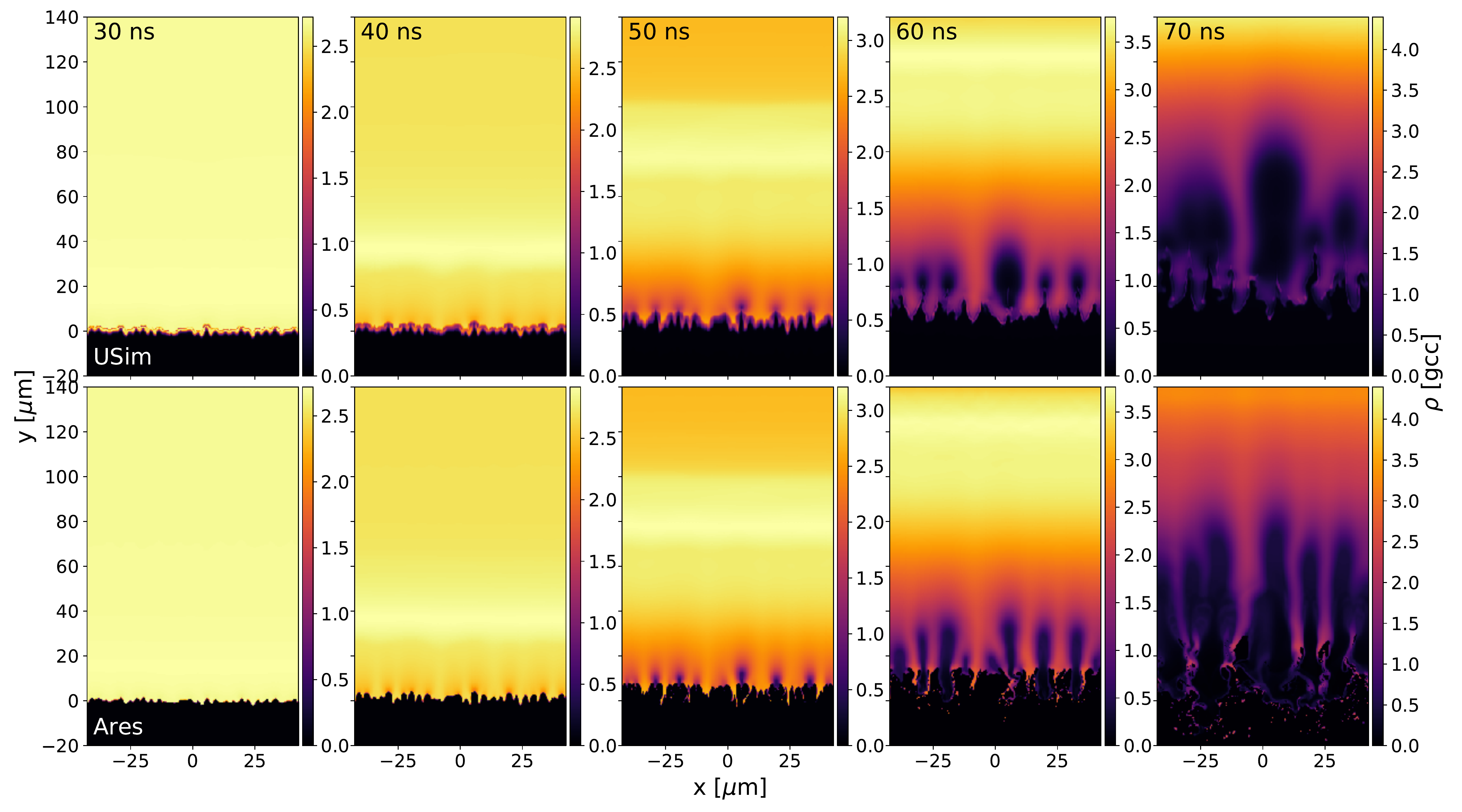}
  \caption{Density plots of the nonlinear ETI simulation outlined in Section~\ref{sec:nonlineti} at different times (correspond to Figure~\ref{fig:rAndI}) with the top row showing the USim results, and the bottom row showing the Ares results. The current at these times is indicated by the green markers in Figure~\ref{fig:rAndI}}\label{fig:ssev}
\end{figure*}

From \citet{slutz2010}, closed-form expressions for liner radius and current are empirically-derived from pulsed-power experiments conducted on the Z machine. The purple dashed curve in Figure~\ref{fig:rAndI} represents the outer liner radius, $r_l$, as a function of time given by
\begin{equation}\label{eqn:rl}
  r_l(t) = r_{l_0} {\left(1-{\left(\frac{t}{t_p}\right)}^4\right)},
\end{equation}
where $r_{l_0}$ is the initial outer liner radius (\SIrange{2.92}{3.168}{\mm}) and $t_p$ is the pulse time ($\approx$\SI{135}{\nano\second}). \cite{slutz2010, sinars2011, peterson2012} Current, $I$, as a function of time is given by 
\begin{equation}\label{eqn:I}
  I(t) = I_x {\left(\frac{27}{4}\right)}^\frac{1}{4}\sqrt{\left(\frac{t}{t_p}\right)^2-{\left(\frac{t}{t_p}\right)}^6},
\end{equation}
where $I_x$ is the peak current (\SIrange{20}{27}{\mega\ampere}). 

Most high power, pulsed-power machines, such as Z, have a prepulse phase before the full pulse is delivered. On the Z machine, this prepulse lasts anywhere from \SIrange{40}{80}{\nano\second} as evident from experimentally-measured load data.\cite{slutz2010, sinars2011, peterson2012} Equation~\ref{eqn:I} does not have slowly rising pre-pulse behavior, which would cause premature ablation disrupting the perturbation of the liner-vacuum interface. To incorporate this experimentally-observed initial rise, the functional form of the current drive in Equation~\ref{eqn:I} is modified to obtain $I_t$. Figure~\ref{fig:rAndI} depicts the original current $I$ and the modified current $I_t$. The adjusted form of the current is
%Specifically Figure~11b of \citet{peterson2012}, represents Z machine measurements of the current, and the time less than \SI{0}{\nano\second} represents the prepulse with slow rise at relatively low currents.
\begin{equation}\label{eqn:It}
  I_t(t) = I(t){\left(1-\exp\left[-{\left(\frac{t}{t_{r}}\right)}^2\right]\right)}^{1.25},
\end{equation}
\rftwo{where $I(t)$ is given by Equation~\ref{eqn:I} and $t_{r}$ is the adjustable pre-pulse time of \SI{60}{\nano\second}}. Note that although the simulations shown in this section are early in time, this current form is usable to accurately approximate late time phenomena in Z-like pulses. Though the peak current of the adjusted form is noticeably smaller than the peak of the original form, it is closer to the experimentally-measured peak current from Figure~11b of \citet{peterson2012}.%Ref.~\onlinecite{peterson2012}.

To convert this current drive to a planar magnetic boundary condition for USim, Ampere's law is used in cylindrical coordinates to determine the magnetic field at the time-varying radial location of the liner interface. For Ares, the current is set as
\begin{equation*}
  I_\text{Ares}(t) = \frac{I_t(t)}{2\pi r_l (t)}, 
\end{equation*}
so that the magnetic field in both planar problems is the same and representative of the magnetic field experienced by the liner on Z. 

\rfone{Ares solves parabolic equations implicitly, contrarily, USim solves parabolic equations semi-explicitly (as discussed in Section~\ref{sec:code}). With an implicit sover, Ares can handle large vacuum resistivity values without excessive computational cost. Because USim uses the semi-explicit STS scheme, it is not computationally practical to run with the same large vacuum resistivity as this leads to an impractically large number of STS stages making the computational cost significantly more expensive.} Thus, a much lower vacuum resistivity needs to be specified with certain constraints. First, if the resistivity of the vacuum is relatively low, this results in unphysical currents in the vacuum, diverting the current away from the liner region. These currents cause unphysical ohmic heating in the vacuum resulting in a highly restrictive time-step. Hence, the vacuum resistivity is set to a large value, and ohmic heating is neglected in the vacuum.\footnote{This violates energy conservation (at least in the vacuum), but is necessary for an accurate current rise in the liner given a finite resistive vacuum when using an explicit or semi-explicit scheme.} Further numerical challenges include the creation of density voids during ETI development due to a finite diffusion rate of the magnetic field through the vacuum. This finite diffusion rate leads to waves in the vacuum creating low density regions where large magnetosonic speeds further restrict the time-step. Avoiding this requires a large enough vacuum resistivity such that the vacuum magnetic diffusion transit time is small relative to the hyperbolic time-step. 

Figure~\ref{fig:simsetup} presents the simulation setup consisting of an $x$ domain of \SI{85}{\micro\metre} and $y$ domain of \SI{250}{\micro\metre} with a resolution of 256x640 cells, respectively. The multimode perturbation of the interface is of the form
\begin{equation} \label{eqn:multipert}
  \delta =\frac{1}{32} \sum_{m=1}^{m=32}\beta_m \text{cos}\left[2\pi\left(\frac{m\cdot x}{\lambda_{max}}+\beta_m\right)\right]\;,
\end{equation}
where $\beta$ is a random number from 0 to 1, $\lambda_{max}=200\;\mu m$ is the maximum wavelength associated with the lowest mode (see \ref{app:coef} for the coefficients used). Figure~\ref{fig:simsetup} shows the initial state. This setup uses a pressure equilibrium to avoid bulk motion of the liner, as material strength models are not employed. Note the EOS interpolation algorithm is the birational LEOS interpolation scheme for Ares\cite{more1988, young1995, fritsch2003} and the birational EOSPAC interpolation scheme for USim\cite{cranfill1983}. For thermal and electrical conductivities of the liner, the QLMD table 29373 for aluminum is used, and the EOS for the vacuum and liner is BMEOS, as discussed in Section~\ref{sec:code:eos}.\cite{desjarlais2002} 

Figure~\ref{fig:ssev} shows snapshots in time of the simulation results for USim (top) and Ares (bottom). \rftwo{Both codes show a similar magnetic diffusion wave and a density spike that is propagating inward in the \SI{40}{\nano\second}, \SI{50}{\nano\second}, and \SI{60}{\nano\second} snapshots at $y\approx\SI{38}{\micro\metre}$, $y\approx\SI{79}{\micro\metre}$, and $y\approx\SI{122}{\micro\metre}$, respectively}.\footnote{Figures~\ref{fig:ssev} uses a subsection of the simulation result and is why the peak is not seen explicitly in the \SI{70}{\nano\second} plot}. While the USim results do not allow ablation into the vacuum, resulting in no mixing of material regions, the Ares results do show ablation into the vacuum, as its multimaterial treatment handles mixing of material regions. \rftwo{This ablation/finger development, seen in the Ares results and slightly in the USim results, is the beginning of the subsequent electro-choric instability (ECI) as discussed by \citet{pecover2015}. Due to the difference in multimaterial treatment between the codes \footnote{USim currently does not have vapourisation capabilities} this is not explored further, but note the time for when ECI begins to form (\SI{60}{\nano\second}) matches the observations from the \citet{pecover2015} simulations.}

\begin{figure}%htb]
  \centering
  \includegraphics[width=1.0\columnwidth]{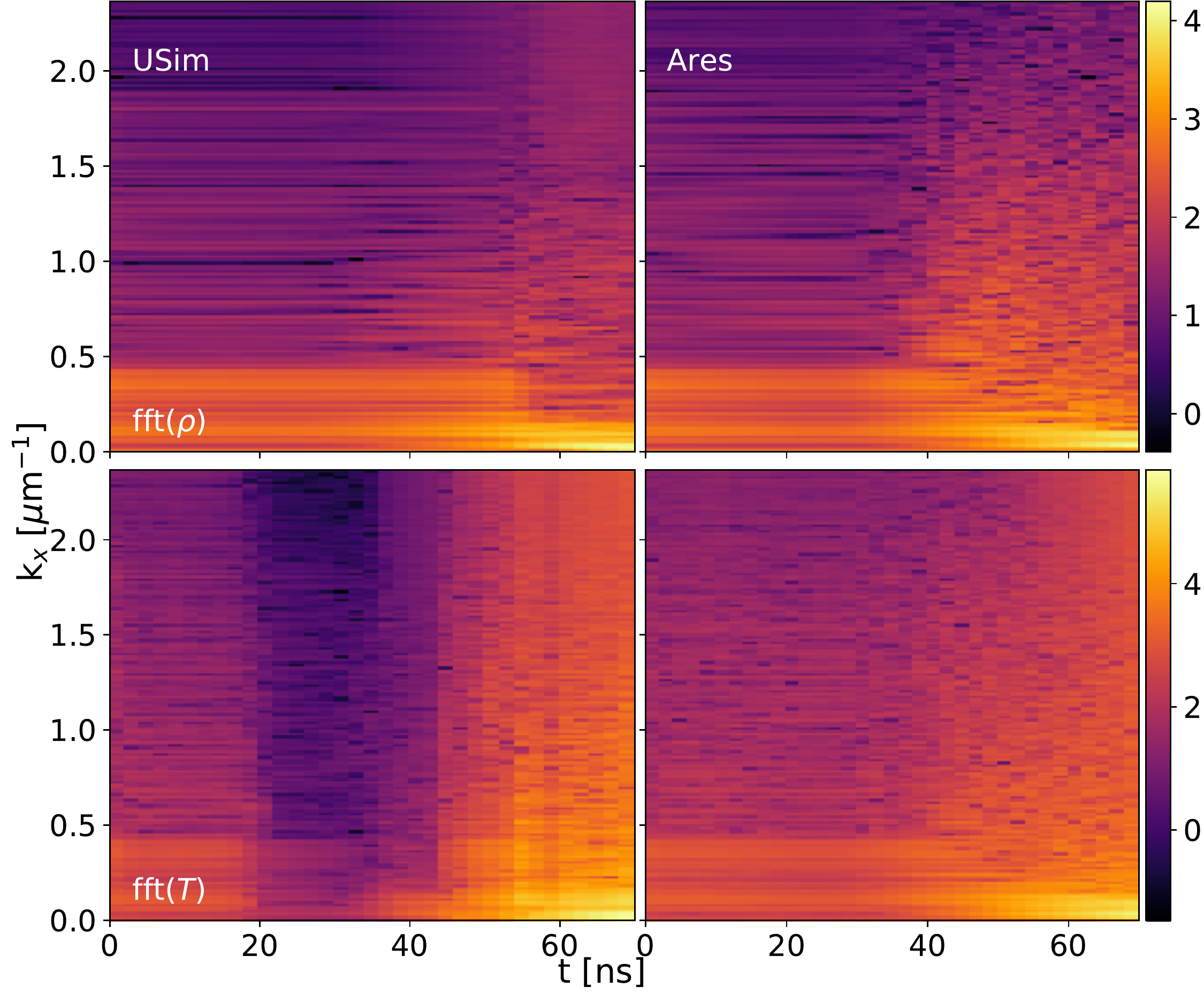}
  \caption{The fast Fourier Transform of y-averaged values, of density (top) and temperature (bottom), over a range of \SIrange{-100}{145}{\micro\metre} along the $x$-direction are presented. Plots shown are different snapshots in time from \SIrange{0}{70}{\nano\second} of the nonlinear ETI simulation outlined in Section~\ref{sec:nonlineti}. The colorbar corresponds to spectral energy, the x-axis corresponds to time, and the y-axis corresponds to the wave number along the $x$-direction.}\label{fig:fft}
\end{figure}

\begin{figure}%htb]
  \centering
  \includegraphics[width=1.0\columnwidth]{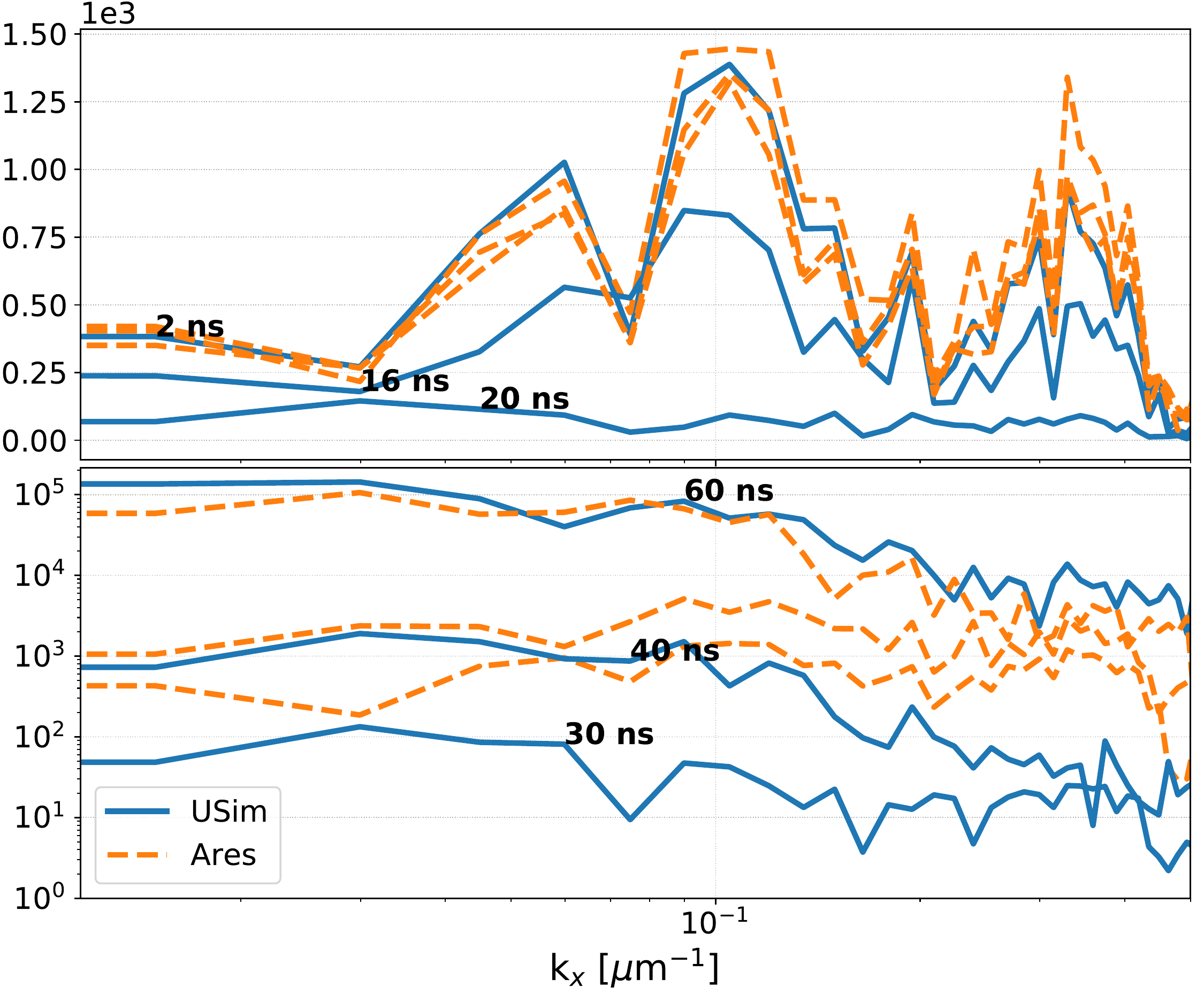}
  \caption{Vertical lineouts of the temperature FFTs is shown in the bottom row of Figure~\ref{fig:fft} at various times for both the Ares and USim results. Note that the differences between the codes are more pronounced early-in-time (top plot) whereas the solutions agree more closely late-in-time (bottom plot).}\label{fig:fft1D}
\end{figure}

Earlier in time, the interface in the USim simulation is diffuse, showing a smoother density gradient relative to the Ares result at \SI{30}{\nano\second} and \SI{40}{\nano\second}. At \SI{70}{\nano\second}, the Ares result shows smaller wavelength growth with a sharper density gradient at the spike interface even though the spikes penetrate to a similar distance ($\approx\SI{40}{\micro\metre}$). At \SI{60}{\nano\second}, both codes show similar wavelength modes of ETI, whereas at \SI{70}{\nano\second}, Ares retains more shorter-wavelength modes compared to USim.  

Figure~\ref{fig:fft} presents a discrete fast Fourier transform of density and temperature along the $x$-direction, and highlights how the mode structures change over time. Figure~\ref{fig:fft} also shows the early time (\SI{30}{\nano\second}, \SI{40}{\nano\second}, and \SI{50}{\nano\second}) suppression of the perturbation in the USim results due to numerical diffusion at the interface. Diffusion is more pronounced in the FFT of temperature (bottom left plot) in Figure~\ref{fig:fft}. Both codes converge later in time to lower mode (lower $k$) growth, as is observed qualitatively in Figure~\ref{fig:ssev} and quantitatively in Figure~\ref{fig:fft}. %Hence, both codes are similar in their late-time low-$k$ behavior while differing in their high-$k$ dynamics. 

Figure~\ref{fig:fft1D} shows FFT of the temperature at several different times corresponding to the decrease and increase in amplitude shown for the USim results in the bottom left plot of Figure~\ref{fig:fft}. By doing so, Figure~\ref{fig:fft1D} highlights the USim result diverging from the Ares result early in time while approaching the Ares result later in time. This suppression and growth of the perturbation in USim could be due to USim's handling of the source terms in the vacuum. This could also be due to the diffusion stencil at the interface reducing mode amplitude early in time, and increases in temperature variation caused by signifcant ohmic heating raising mode amplitude late in time. 

\subsection{EOS Sensitivity}\label{sec:nonlineti:eos}

\begin{figure}%htb]
  \centering
  \includegraphics[width=1.0\columnwidth]{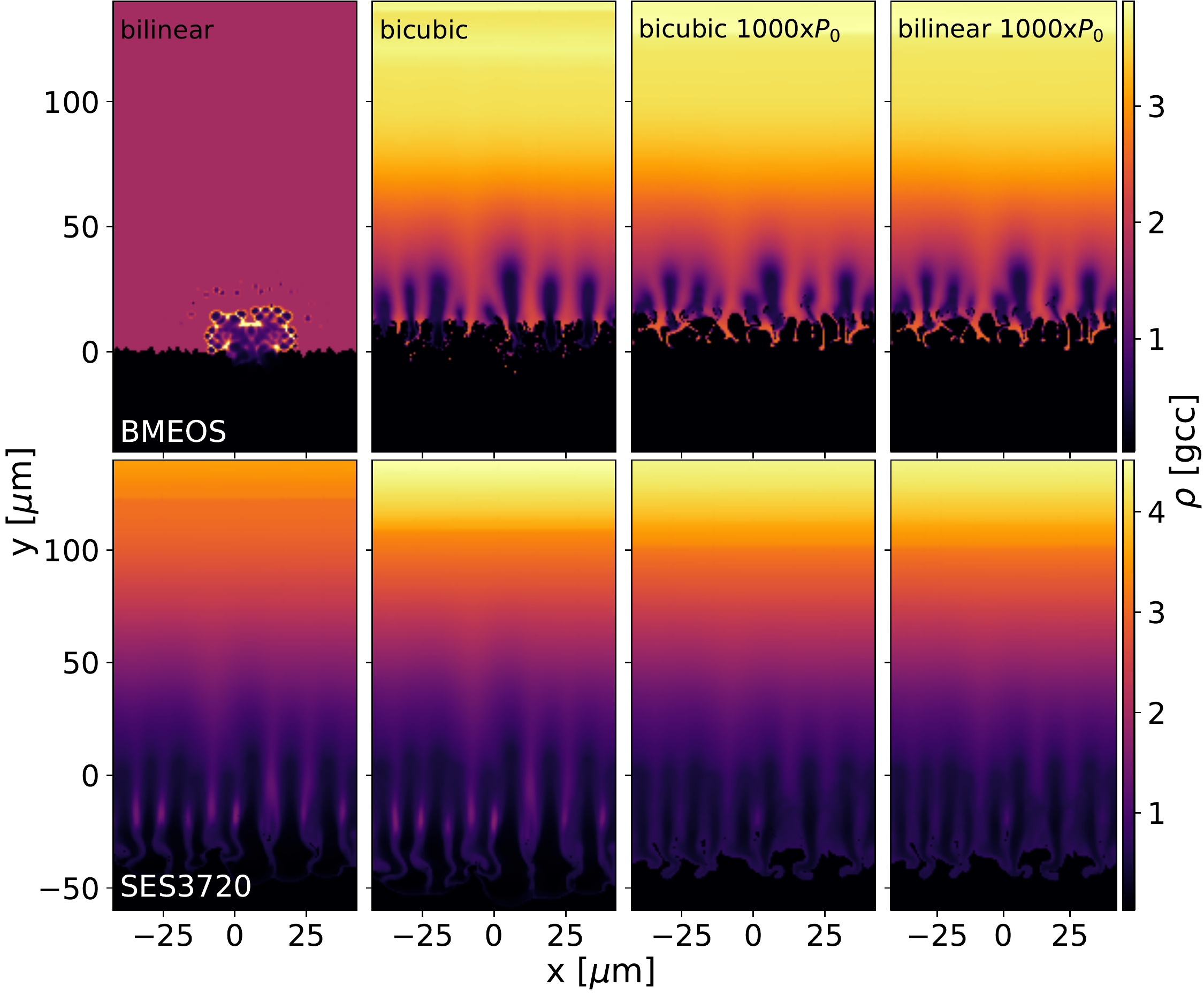}
  \caption{Density evolution of the nonlinear ETI simulations is presented at \SI{60}{\nano\second} using different EOS interpolation schemes (bilinear and bicubic) and different EOS (BMEOS and SES3720).\cite{SESAME3720}}\label{fig:eos}
\end{figure}

A sensitivity analysis performed with Ares shows how different choices of EOS and interpolation scheme impact the development of ETI in the nonlinear regime. From Equation~\ref{eqn:etigr}, the growth rate of ETI is dependent on $\epsilon_T$ (specific heat capacity), and this value is deduced from the representative EOS. Choices of table interpolation, inversion, and monotonicity, will influence the nonlinear ETI behavior directly through $\epsilon_T$ (indirectly through $\eta_T$). 

Figure~\ref{fig:eos} shows the effect of using different EOS on the ETI mode growth in the Ares simulations with the BMEOS (top) and the SES3720 EOS (bottom). Using bilinear interpolation at low pressure and high density results in the crash of both the BMEOS and the SES3720 simulations as noted in the leftmost subplots (top for BMEOS and bottom for SES3720) of Figure~\ref{fig:eos}. \rfone{This crash is from the evaluation of the sound speed, obtained from derivatives of the pressure with respect to density and temperature, encountering imaginary values (imaginary time-step) for both EOS simulations}. \rftwo{The qualitative differences between the bilinear interpolation BMEOS and SES3720 cases before the simulations crash (leftmost plots) are likely due to the resolution of the table at the high-density, low-temperature regime and/or the magnitude of the sound speed evaluated in this region.} Using bicubic interpolation produces the same result as more high fidelity interpolation schemes such as birational, bihermite, bimonotonic, and biquintic, not shown here.   %(see \citet{king2020} for more details on tabulated sound speed reconstruction)

The bilinear interpolation becomes suitable in the high-pressure, high-density regime, as this is a better-defined region of the table.\footnote{Better as in satisfying monotonicity, or having positive pressure and energy values. Negative values result from the imaginary state space of the Van Der Waal's isotherm loops} The two right-most columns of Figure~\ref{fig:eos} reflect this by showing no difference between bicubic and bilinear interpolation, and the simulation is able to run to completion. Increasing the initial pressure moves the initial state to higher temperature (not linearly) and into a smoother region in both tables, permitting the simulation to run to completion. This smoother region has a 30 percent difference between the BMEOS and SES3720, while the lower initial pressure region has 40 percent difference. This difference is still large and explains the qualitatively different result in the 2 right-most columns of Figure~\ref{fig:eos}. Initial pressure is varied instead of density because the electrical resistivity is highly sensitive to density near reference solid density, as mentioned in \citet{peterson2012}, and it would change the $\eta_T$ value through differences in collisional quanities.

Qualitatively the representative SES3720 and BMEOS simulations show differences late in time, while early in time they show similar smaller wavelength mode structure (not shown here). These differences late in time highlight the importance of the EOS on nonlinear ETI development. For the parameters surveyed in this study, the dependence on interpolation algorithm is not as significant as long as an interpolation scheme is used with higher fidelity than bilinear interpolation. 

%Although it was not explored here there is significance on the inversion process used especially when near poorly defined regions of the table as this can lead to different temperatures inverted from the EOS pressure table.

%The effect of interpolation on ETI growth is shown in Figure~\ref{fig:eos} where bilinear interpolation fails to run to completion failing before hand. The middle three plots using bicubic, bihermite, and bimonotonic, respectively, show no significant difference in the end snapshot relative to the Ares birational result on the bottom right of Figure~\ref{fig:ssev}. From this any interpolation scheme with higher fidelity than bilinear interpolation should be adequate to capture ETI growth. Although qualitatively bicubic, bihermite, bimonotonic, and birational are similar in Figure~\ref{fig:eos} the number of cycles and timestep restriction is different. The timestep restriction is based on the soundspeed which requires thermodynamic derivatives of the EOS outlined in more detail ({\color{red}REF EOS Paper}). 

\subsection{Vacuum Resistivity Sensitivity}\label{sec:nonlineti:res}

%\rftwo{In the nonlinear ETI simulation setup the vacuum resistivity is uniformly set at the initial vacuum resistivity of \SI{5.65e-4}{\ohm\metre} for the entire simulation}. 

\rfone{In this work, the vacuum is treated as a separate material from the liner that has a fixed (i.e., constant) large resistivity (\SI{5.65e-5}{\ohm\metre}) while the liner material uses tabulated electrical conductivity (see Section~\ref{sec:code:eos} and Figure~\ref{fig:simsetup}). The only difference between the treatment of the liner and vacuum regions in these Ares simulations is in the different electrical and thermal conductivities. The mesh is initialized such that the interface contains multi-material zones which are handled with the method stated in Section~\ref{sec:code:ares}. The presence of multiple-materials due to the liner being ejected into the vacuum is evident in the late-stage evolution presented in Section \ref{sec:nonlineti:base} (\SI{60}{\nano\second} and \SI{70}{\nano\second})}.

%In the nonlinear ETI study, the problem is initialized with the mesh conformal to the liner interface (IS THIS TRUE??) so there are no multimaterial zones at t=0

\begin{figure}%htb]
  \centering
  \includegraphics[width=1.0\columnwidth]{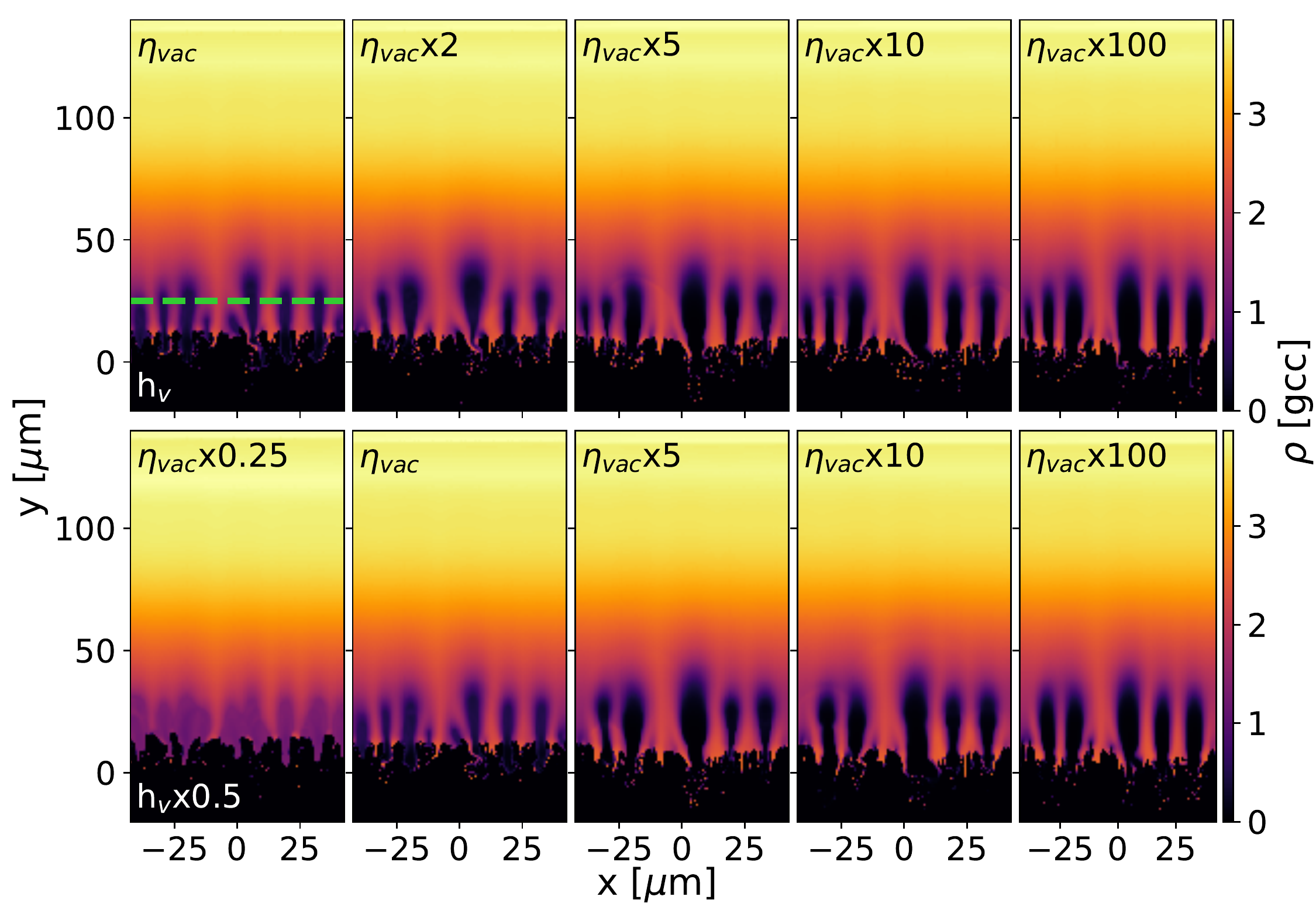}
  \caption{\rftwo{Density evolution of the nonlinear ETI is presented at \SI{60}{\nano\second} with the resistivity varying by column and the initial height of the vacuum ($h_v$) varying by row. The bottom row is for half of the vacuum height compared to the top row. The green dashed line indicates the lineout location used for the top row to generate Figure~\ref{fig:resiLO}.}}\label{fig:resi}
\end{figure}

The vacuum resistivity changes the rate that the magnetic field diffuses through the vacuum and is important in evaluating the ohmic heating due to the spatial variation in resistivity. For explicit codes, the vacuum resistivity should be as low as possible while still achieving a similar result to the infinitely resistive vacuum limit. To determine the role of vacuum resistivity on ETI development, the Ares nonlinear ETI simulations are repeated while multiplying the vacuum resistivity by up to 100 times the nominal value used in the preceding studies ($\SI{5.65e-5}{\ohm\metre}$). 

\begin{figure}%htb]
  \centering
  \includegraphics[width=0.9\columnwidth]{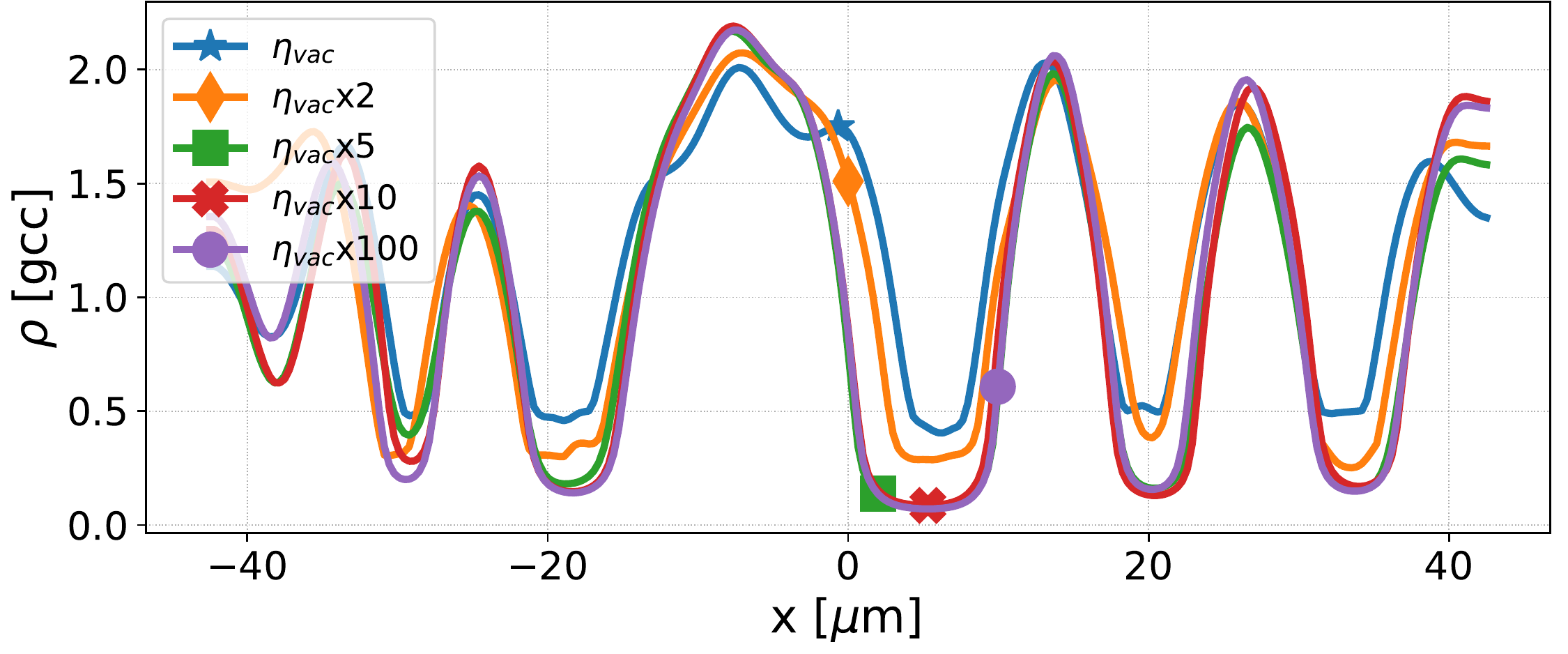}
  \caption{\rftwo{Horizontal lineouts of the top row of Figure~\ref{fig:resi} is presented for all vacuum resistivity cases with the original vacuum height ($h_v$). The horizontal lineout, indicated by the dashed green line in the upper leftmost plot of Figure~\ref{fig:resi}, is at $y=\SI{25}{\micro\metre}$ for all cases.}}\label{fig:resiLO}
\end{figure}

Figure~\ref{fig:resi} shows the result of varying the resistivity and height of the vacuum. As the resistivity increases in the vacuum, the solution converges as noted in the upper row of Figure~\ref{fig:resi}. Qualitatively, the 5x, 10x, and 100x $\eta_v$ simulations show no differences, but the 1x and 2x $\eta_v$ simulations show noticeable differences. \rftwo{The main difference is in the magnitude of density inside the liner hotspots. Figure~\ref{fig:resiLO} presents horizontal lineouts at $y=\SI{25}{\micro\metre}$ for the top row of Figure~\ref{fig:resi}. These lineouts show that the magnitude of density inside the liner hotspots varies greatly (50\% at $x=\SI{5}{\micro\metre}$) between the 1x and 2x runs, and varies little ($<$5\% at $x=\SI{5}{\micro\metre}$) between the 5x, 10x, and 100x runs}. \rfthree{These results indicate that a resistivity ratio of approximately \SI{5.65e-4}{\ohm\metre} (i.e., 10x, which has $<$1\% variation with the 100x run and the not-shown 1000x run) is sufficient in capturing ETI in the infinitely resistive limit for this nonlinear ETI setup.}
%\rftwo{One difference is in the pinching near the interface of the internal liner hotspots seen in the 1x and 2x runs, but not in the 5x, 10x, and 100x runs. Another difference is in the magnitude of density inside the liner hotspots which show large variaiton (50\%) between the 1x and 2x runs, and little variation ($<$5\%) between the 5x, 10x, and 100x runs}. 

  To investigate the underlying mechanism for the converged vacuum resistivity threshold, an additional resistivity scan is performed in a configuration with half the original vacuum height. \rfone{The vacuum height is the initial spatial distance in the y-direction between the liner-vacuum interface and the outer edge of the vacuum region (lower boundary in the y-direction).  For the simulations presented in Figure~\ref{fig:resi}, the vacuum height, $h_v$, is \SI{100}{\micro\metre}}. The bottom row of Figure~\ref{fig:resi} shows the reduced vacuum size simulations of varying vacuum resistivity. Reducing the vacuum height probes the influence of the characteristic time for the magnetic field to diffuse through the vacuum on ETI development. Reducing the vacuum height by a half would result in a quarter of the vacuum resistivity needed to maintain the same vacuum magnetic diffusion transit time. Reducing the vacuum resistivity needed for a converged result would be beneficial for codes with explicit diffusion algorithms. 

  The two left-most plots of Figure~\ref{fig:resi} have the same magnetic diffusion transit time, but show starkly different ETI growth. This implies the vacuum magnetic diffusion transit time is not an underlying mechanism for the converged vacuum resistivity threshold, suprisingly. The converged vacuum resistivity is approximately 5 times the nominal resistivity from previous studies, and is the same for both vacuum sizes. A more relevant scale parameter for the converged results may be the ratio of the liner resistivity to the vacuum resistivity, as this influences the reconstruction of derivatives given the spatially-varying resistivity. Changing the numerical derivatives leads to different values of current through its impact on ohmic heating, thereby producing differences in the nonlinear ETI growth.   
  
\rfone{Based on these findings, to get the converged ETI result (infinitely resistive vacuum) requires a minimum vacuum-to-liner resistivity ratio of \SI{2e4}{}}. Ares is used here because of the challenges associated with performing such a convergence study with an explicit diffusion vacuum model, such as with USim's STS diffusion algorithm, so care is needed when checking the convergence of vacuum resistivity using explicit codes.

\subsection{Vacuum Density Sensitivity}\label{sec:nonlineti:den}

When simulating a vacuum using a fluid code, the vacuum density is traditionally set relatively low. \cite{peterson2012, peterson2013} Not evolving ohmic heating, $S_\epsilon$ in Equation~\ref{eqn:ener}, and the magnetic acceleration of the vacuum, Equation~\ref{eqn:mtm} through $S_\mathbf{B}$ in Equation~\ref{eqn:magf}, allows for a less restrictive time step by reducing the sound speed. \rfone{For the nonlinear ETI simulations, the vacuum density is evolved with a floor value of the initial vacuum density (\SI{2.7e-2}{\gram\per\cubic\centi\metre}). Varying this floor value determines the effect of vacuum hydrodynamics on nonlinear ETI behavior for this particular setup.}

\begin{figure}%htb]
  \centering
  \includegraphics[width=1.0\columnwidth]{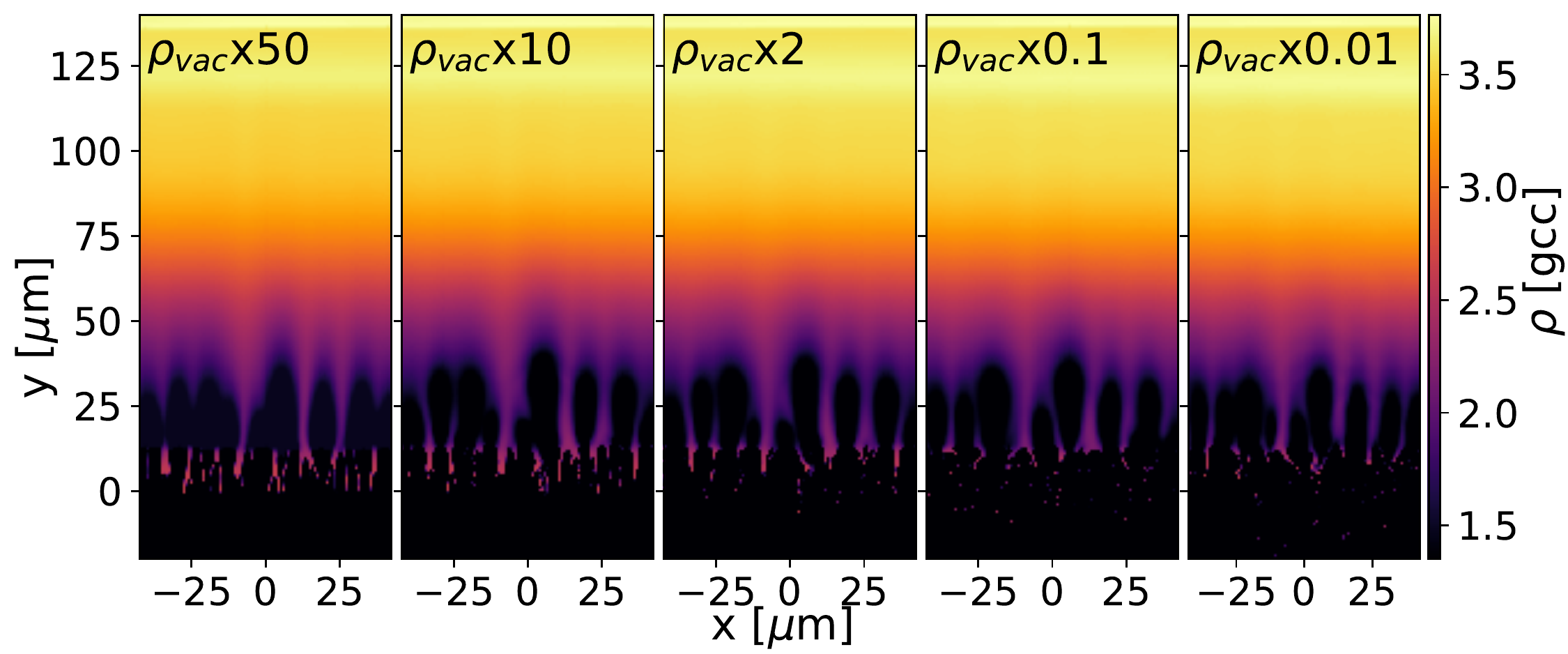}
  \caption{Density evolution of the nonlinear ETI simulations is presented at \SI{60}{\nano\second} varying the vacuum density.}\label{fig:dens}
\end{figure}

\rftwo{Figure~\ref{fig:dens} shows the nonlinear ETI simulation at \SI{60}{\nano\second} for vacuum density varying from 50x to 0.01x of the base vacuum density ($\SI{2.7e-5}{\kilo\gram\per\cubic\metre}$)}. There is no discernable difference between the simulations in Figure~\ref{fig:dens}. All other values of vacuum density show qualitatively similar results when neglecting ohmic heating and magnetic acceleration of the vacuum. Running with too small of a vacuum density leads to long simulation times due to short time steps required to resolve potential hydrodynamic (acoustic) oscillations in the vacuum. More moderate values of vacuum density of approximately 2 to 3 orders of magnitude lower than the reference liner density produce converged results. Using such a large vacuum density only changes the dynamics when significant vacuum inertia is added that impedes the ablation of the liner. Including magnetic acceleration and ohmic heating would change the outcome of this converged density ratio, but is not explored here. 

\section{Conclusion}\label{sec:conclusion}

This work compares nonlinear ETI simulations using two different codes with significantly different algorithmic approaches to solving the resistive-MHD equations. Although these codes differ in many ways, the most significant difference, relative to simulating nonlinear ETI, is in the spatial and temporal discretization of the diffusion terms as discussed in Section~\ref{sec:code}. The handling of these terms directly affects the evaluation of the magnetic diffusion wave and ohmic heating which are both essential in simulating nonlinear ETI growth. Furthermore, the range of viable parameters for stable and efficient computational results is dictated by the discretization methods. For these nonlinear ETI simulation comparisons, a tabulated EOS for aluminum, BMEOS, is developed that compares well with the previously used SES3720 table. \cite{peterson2012}

Section~\ref{sec:comp} shows development of verification test cases evaluating each code's diffusion capabilities for $S_\epsilon$ and $S_\mathbf{B}$ from Equations~\ref{eqn:ener}~and~\ref{eqn:magf}, respectively. First, the codes simulate a simple magnetic diffusion test case showing that both codes accurately recover the analytical solution. Next, the codes recover the theoretical linear ETI growth rate shown in Equation~\ref{eqn:sgr}. Comparing the simulation growth rate to the theoretical one for varying spatial resolution and time step size results in the convergence shown in Figure~\ref{fig:etigrloglog}, where the anticipated orders of accuracy of spatial and temporal discretization are obtained. These tests give confidence in each code's ability to handle the fundamental aspects of simulating nonlinear ETI. 

Section~\ref{sec:nonlineti} compares the codes for simulating nonlinear ETI with the baseline setup shown in Figure~\ref{fig:simsetup}. This simulation has full coupling of both source terms $S_\epsilon$ and $S_\mathbf{B}$ to the full set of MHD equations, Equations~\ref{eqn:cont}-\ref{eqn:magf}, and also includes the non-ideal-gas BMEOS. The simulation uses an analytic form for the current rise akin to a typical Z-machine current rise with a correct prepulse as represented in Figure~\ref{fig:rAndI}. Figures~\ref{fig:ssev}-\ref{fig:fft1D} show qualitative and quantitative agreement between the two codes, although interesting differences arise in the details of mode evolution. 

Additionally, the Ares nonlinear ETI simulation undergoes a sensitivity analysis for the vacuum conditions shown in Sections~\ref{sec:nonlineti:res}~and~\ref{sec:nonlineti:den}. The analyses show a strong dependence on the vacuum resistivity, but not the vacuum density. Specifically, the resistivity analysis shows that a vacuum-to-liner resistivity ratio of $\approx\SI{2e4}{}$ is sufficient to capture the converged (i.e., the infinitely resistive vacuum limit) nonlinear ETI simulation. The EOS implementation is tested across different interpolation algorithms, EOS tables, and initial conditions, showing a large dependence of nonlinear ETI growth to the EOS table and interpolation algorithm, specifically near the solid density and low pressure state. These sensitivity analyses provide guidelines for how codes that explicitly integrate diffusion terms can still capture nonlinear ETI without the need for a infinitesimally dense and infinitely resistive vacuum.  

\section*{Acknowledgement}
This work was supported through the Lawrence Livermore National Laboratory Weapons and Complex Integration (LLNL WCI) High Energy Density Fellowship, and through the US Department of Energy under grants DE-SC0016515, DE-SC0016531, \& DE-NA0003881. In addition a portion of this work was sponsored by LLNL WCI HED summer program. This research used resources of the National Energy Research Scientific Computing Center (NERSC), a U.S. Department of Energy Office of Science User Facility operated under Contract No. DE-AC02-05CH11231. This document has been approved for release under LLNL-JRNL-812448. 

A portion of this work was performed under the auspices of the U.S. Department of Energy by Lawrence Livermore National Laboratory under Contract DE-AC52-07NA27344. This document was prepared as an account of work sponsored by an agency of the United States government. Neither the United States government nor Lawrence Livermore National Security, LLC, nor any of their employees makes any warranty, expressed or implied, or assumes any legal liability or responsibility for the accuracy, completeness, or usefulness of any information, apparatus, product, or process disclosed, or represents that its use would not infringe privately owned rights. Reference herein to any specific commercial product, process, or service by trade name, trademark, manufacturer, or otherwise does not necessarily constitute or imply its endorsement, recommendation, or favoring by the United States government or Lawrence Livermore National Security, LLC. The views and opinions of authors expressed herein do not necessarily state or reflect those of the United States government or Lawrence Livermore National Security, LLC, and shall not be used for advertising or product endorsement purposes.

\bibliographystyle{elsarticle-num-names}
\bibliography{reference}

\appendix

\section{Comparison of BMEOS and SES3720}\label{app:bmeos}

\begin{figure}[H]
  \includegraphics[width=1.0\columnwidth]{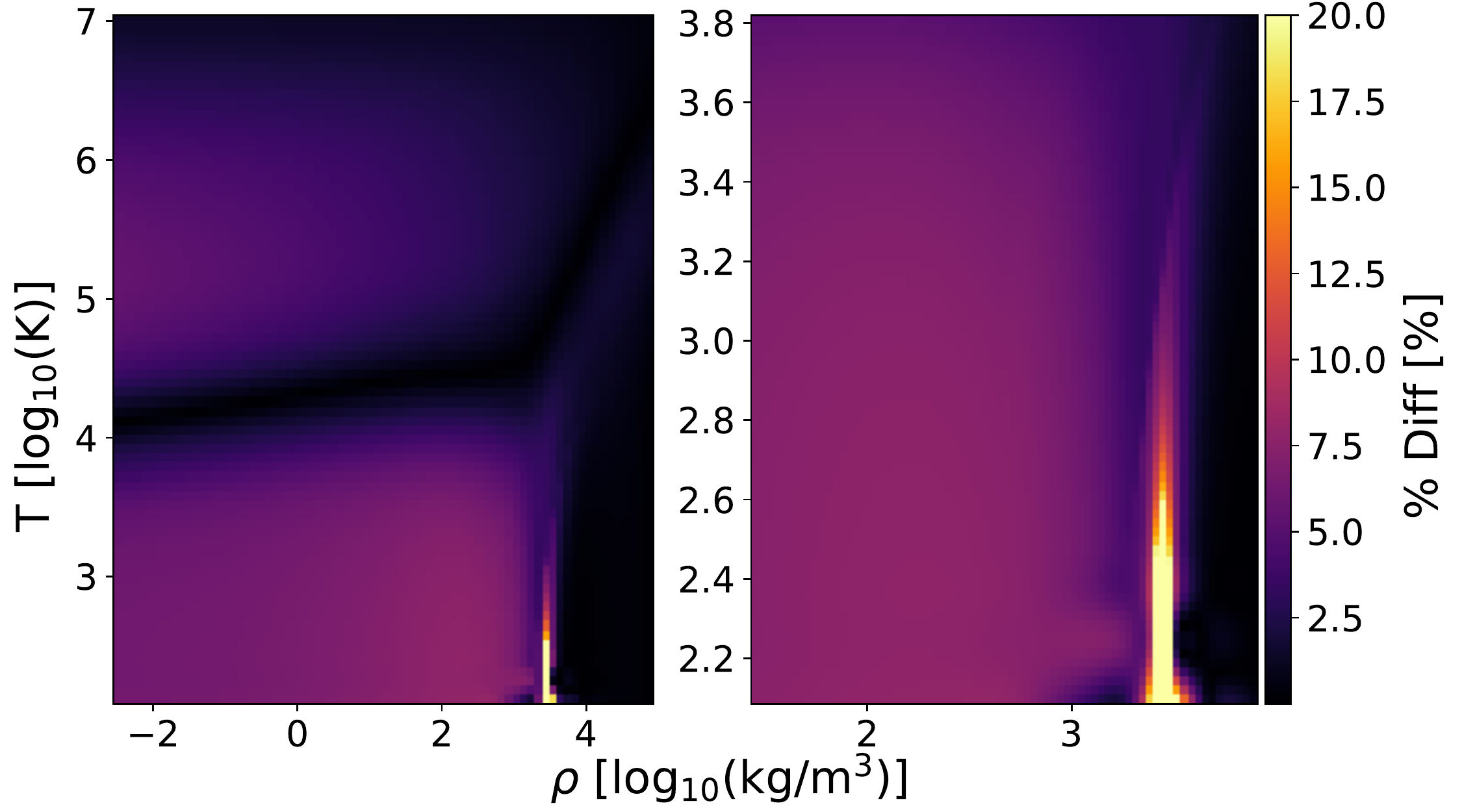}
  \caption{Percent difference of the specific internal energy density between the BMEOS and SES3720 in the state space relevant to the nonlinear ETI evolution in HED regimes. The plot on the right is an expanded scale of the left plot to highlight the region of largest difference.} \label{fig:eosdiff}
\end{figure}

\section{Coefficients of the multimode perturbation}\label{app:coef}

\begin{table}[htbp]
  \begin{tabular} {|c|c|}
    \hline
    mode (i) & $\beta_i$ \\
    \hline
    1  &  0.883494  \\  %88349446717067630
    2  &  0.313251  \\  %31325129550977693
    3  &  0.139670  \\  %13967043719620320
    4  &  0.438109  \\  %43810854286921563
    5  &  0.642904  \\  %64290417015051950
    6  &  0.176107  \\  %17610703584453546
    7  &  0.856669  \\  %85666854117495270
    8  &  0.630685  \\  %63068509642117250
    9  &  0.682887  \\  %68288697221320270
    10 &  0.941226  \\  %94122571696996590
    11 &  0.236611  \\  %23661083309690045
    12 &  0.699510  \\  %69951021548430460
    13 &  0.440243  \\  %44024323864935710
    14 &  0.124690  \\  %12468958135598729
    15 &  0.643533  \\  %64353305518563500
    16 &  0.018313  \\  %01831296188057540
    17 &  0.415389  \\  %41538878723351090
    18 &  0.403712  \\  %40371217000075044
    19 &  0.122180  \\  %12217984072116572
    20 &  0.313884  \\  %31388390115349750
    21 &  0.207358  \\  %20735848022418413
    22 &  0.915150  \\  %91514973969305620
    23 &  0.038463  \\  %03846332536563790
    24 &  0.991615  \\  %99161494860361870
    25 &  0.755673  \\  %75567300310861700
    26 &  0.558353  \\  %55835324108543100
    27 &  0.586421  \\  %58642051221038750
    28 &  0.896183  \\  %89618281121054160
    29 &  0.305981  \\  %30598100465018050
    30 &  0.495188  \\  %49518773592475260
    31 &  0.476349  \\  %47634898809425210
    32 &  0.057556  \\  %05755578148740470
    \hline
  \end{tabular}
\end{table}

\end{document}